\documentclass[a4paper,UKenglish,cleveref,autoref,thm-restate]{lipics-v2021}

\def\fullversion

\usepackage{etoolbox}

\usepackage{booktabs}   
\usepackage{subcaption} 
\usepackage{caption}
\usepackage{hhline}
\usepackage{colortbl}
\usepackage{style}
\usepackage{bigstrut}
\usepackage{microtype}
\usepackage{tabularx}
\usepackage{multirow}
\usepackage{floatrow}
\usepackage{multicol}
\usepackage{rotating}
\usepackage{lipsum}
\usepackage{balance}
\usepackage{mfirstuc}
\usepackage{titlecaps}
\usepackage{listings}
\usepackage{float}
\usepackage[normalem]{ulem}
\usepackage{adjustbox}

\usepackage{xcolor}

  \newcommand{\codeskip}{{\vspace{.05in}}}
  
  \newcommand{\fname}[1]{\textit{#1}}

  \newcommand{\true}{\emph{true}\xspace}
  \newcommand{\false}{\emph{false}\xspace}

  \newcommand{\polylog}{\mathrm{polylog}}

  \newcommand{\hashfunc}{h}
  \newcommand{\hashcombine}{\oplus}
  \newcommand{\hashdiff}{\ominus}
  \newcommand{\mmod}{~\text{mod}~}
  \newcommand{\prefixhash}[1]{\mathcal{P}_{#1}}
  \newcommand{\suffixhash}[1]{\mathcal{S}_{#1}}
  \newcommand{\remain}[1]{\mathit{rest}_{#1}}
  \newcommand{\nextblock}[1]{r_{#1}}

  \newcommand{\bsize}{b}
  
  \newcommand{\dpv}{G}
  \newcommand{\monge}{Monge}
  \newcommand{\dist}{\mathit{D}}
  \newcommand{\prefix}{\mathit{prefix}}
  \newcommand{\seqA}{A}
  \newcommand{\seqB}{B}
  \newcommand{\odd}{odd}
  \newcommand{\even}{even}
  \newcommand{\SP}{SD}
  \newcommand{\shortestdis}{shortest distance}

  \newcommand{\algosty}[1]{\textsc{#1}}
  \newcommand{\bfssa}{\algosty{BFS-SA}\xspace}
  \newcommand{\bfssimplehash}{\algosty{BFS-Hash}\xspace}
  \newcommand{\bfshash}{\algosty{BFS-B-Hash}\xspace}
  \newcommand{\dacmm}{\algosty{DaC-SD}\xspace}
  \newcommand{\dacmmrec}{\algosty{GetDistance}\xspace}
  \newcommand{\hashtabfull}{prefix table\xspace}
  
  \newcommand{\hashtab}{prefix table\xspace}
  \newcommand{\aalm}{AALM\xspace}

  \newcommand{\block}{block\xspace}

  \newcommand{\LCP}{LCP}

  \newcommand{\dx}{\mathit{dx}}
  \newcommand{\dy}{\mathit{dy}}

  \newcommand{\longestprefix}{\fname{LCP}}

  \newcommand{\forkins}{\texttt{fork}}

  \newcommand{\thread}{thread}

\SetCommentSty{mycommfont}
\SetKwFor{parForEach}{parallel-for-each}{do}{endfor}
\SetKwProg{MyStruct}{Struct}{}{end}
\SetKwProg{MyFunc}{Function}{}{end}
\newcommand{\nosemic}{\renewcommand{\@endalgocfline}{\relax}}
\newcommand{\dosemic}{\renewcommand{\@endalgocfline}{\algocf@endline}}
\SetSideCommentLeft

\crefname{section}{Sec.}{Sec.}
\crefname{algocf}{Alg.}{Algs.}
\Crefname{algocf}{Alg.}{Algs.}
\crefname{theorem}{Thm.}{Thm.}
\crefname{table}{Tab.}{Tab.}
\crefname{figure}{Fig.}{Fig.}

\title{Efficient Parallel Output-Sensitive Edit Distance} 

\titlerunning{Efficient Parallel Output-Sensitive Edit Distance} 

\author{Xiangyun Ding}{University of California, Riverside, CA, USA}{xding047@ucr.edu}{https://orcid.org/0009-0001-8367-8399}{}
\author{Xiaojun Dong}{University of California, Riverside, CA, USA}{xdong038@ucr.edu}{https://orcid.org/0000-0003-4828-7066}{}
\author{Yan Gu}{University of California, Riverside, CA, USA}{ygu@cs.ucr.edu}{https://orcid.org/0000-0002-4392-4022}{}
\author{Youzhe Liu}{University of California, Riverside, CA, USA}{yliu908@ucr.edu}{https://orcid.org/0009-0004-9721-5522}{}
\author{Yihan Sun}{University of California, Riverside, CA, USA}{yihans@cs.ucr.edu}{https://orcid.org/0000-0002-3212-0934}{}

\authorrunning{X. Ding, X. Dong, Y. Gu, Y. Liu, and Y. Sun} 

\Copyright{Xiangyun Ding, Xiaojun Dong, Yan Gu, Youzhe Liu, Yihan Sun} 



\category{} 

\newcommand{\ifconference}[1]{{{\ifx\fullversion\undefined{#1}\fi}}}
\newcommand{\iffullversion}[1]{{{\ifx\conference\undefined{#1}\fi}}}

\ifconference{
\relatedversiondetails[cite=ding2023efficientfull]{Full Version}{https://arxiv.org/abs/2306.17461} 
}

\supplement{}
\supplementdetails[subcategory={Source Code}, cite={edcode}]{Software}{https://github.com/ucrparlay/Edit-Distance} 

\funding{This work is supported by NSF grants CCF-2103483, CCF-2238358, and IIS-2227669, and UCR Regents Faculty Fellowships.}

\EventEditors{Inge Li G{\o}rtz, Martin Farach-Colton, Simon J. Puglisi, and Grzegorz Herman}
\EventNoEds{4}
\EventLongTitle{31st Annual European Symposium on Algorithms (ESA 2023)}
\EventShortTitle{ESA 2023}
\EventAcronym{ESA}
\EventYear{2023}
\EventDate{September 4--6, 2023}
\EventLocation{Amsterdam, the Netherlands}
\EventLogo{}
\SeriesVolume{274}
\ArticleNo{46}

\makeatletter
\patchcmd{\@algocf@start}
  {-1.5em}
  {0pt}
  {}{}
\makeatother

\hideLIPIcs{}
\nolinenumbers

\begin{document}

\maketitle
\begin{abstract}
In this paper, we study efficient parallel edit distance algorithms, both in theory and in practice.
Given two strings $A[1..n]$ and $B[1..m]$, and a set of operations allowed to edit the strings,
the edit distance between $A$ and $B$ is the minimum number of operations required to transform $A$ into $B$.
In this paper, we use edit distance to refer to the Levenshtein distance, which allows for unit-cost single-character edits (insertions, deletions, substitutions).
Sequentially, a standard Dynamic Programming (DP) algorithm solves edit distance with $\Theta(nm)$ cost.
In many real-world applications, the strings to be compared are similar to each other and have small edit distances.
To achieve highly practical implementations, we focus on output-sensitive parallel edit-distance algorithms,
i.e., to achieve asymptotically better cost bounds than the standard
$\Theta(nm)$ algorithm when the edit distance is small.
We study four algorithms in the paper, including three algorithms based on Breadth-First Search (BFS), and one algorithm based on Divide-and-Conquer (DaC).
Our BFS-based solution is based on the Landau-Vishkin algorithm.
We implement three different data structures for the longest common prefix (LCP) queries needed in the algorithm:
the classic solution using parallel suffix array, and two hash-based solutions proposed in this paper.
Our DaC-based solution is inspired by the output-insensitive solution proposed by Apostolico et al.,
and we propose a non-trivial adaption to make it output-sensitive.
All of the algorithms studied in this paper have good theoretical guarantees, and
they achieve different tradeoffs between work (total number of operations), span (longest dependence chain in the computation),
and space.

We test and compare our algorithms on both synthetic data and real-world data, including DNA sequences, Wikipedia texts, GitHub repositories, etc.
Our BFS-based algorithms outperform the existing parallel edit-distance implementation in ParlayLib in all test cases.
On cases with fewer than $10^5$ edits, our algorithm can process input sequences of size $10^9$ in about ten seconds,
while ParlayLib can only process sequences of sizes up to $10^6$ in the same amount of time.
By comparing our algorithms, we also provide a better understanding of the choice of algorithms for different input patterns.
We believe that our paper is the first systematic study in the theory and practice of parallel edit distance.
\end{abstract}

\newpage

\section{Introduction}
\label{sec:intro}


Given two strings (sequences) $A[1..n]$ and $B[1..m]$ over an alphabet $\Sigma$ and a set of operations allowed to edit the strings,
the \defn{edit distance} between $A$ and $B$ is the minimum number of operations required to transform $A$ into $B$.
WLOG, we assume $m\le n$.
The most commonly used metric is the \defn{Levenshtein distance} which allows for unit-cost single-character edits (insertions, deletions, substitutions).
In this paper, we use \emph{edit distance} to refer to the Levenshtein distance. 
We use $k$ to denote the edit distance for strings $A$ and $B$ throughout this paper.
Edit distance is usually used to measure the similarity of two strings (a smaller distance means higher similarity).

Edit distance is a fundamental problem in computer science, and is introduced in most algorithm textbooks (e.g.,~\cite{CLRS,dasgupta2008algorithms,goodrich2015algorithm}).
In practice, it is widely used in version-control software~\cite{spinellis2012git}, computational biology~\cite{cheon2015homomorphic,jiang2002general,li2010survey}, natural language processing~\cite{boucher2022bad,hossain2019auto}, and spell corrections~\cite{hladek2020survey}.
It is also closely related to other important problems such as
longest common subsequence (LCS)~\cite{paterson1994longest}, longest increasing subsequence (LIS)~\cite{krusche2010new}, approximate string matching~\cite{ukkonen1985algorithms}, and multi-sequence alignment~\cite{zhang2003alignment}.
The classic dynamic programming (DP) solution can compute edit distance in $O(nm)$ work (number of operations) between two strings of sizes~$n$ and~$m$.
This complexity is impractical if the input strings are large.
One useful observation is that, in real-world applications, the strings to be compared are usually \emph{reasonably similar}, resulting in a relatively small edit distance.
For example, in many version-control softwares (e.g., Git), if the two committed versions are similar (within a certain number of edits),
the ``delta'' file is stored to track edits.
Otherwise, if the difference is large, the system directly stores the new version.
Most of the DNA or genome sequence alignment applications also only focus on when the number of edits is \emph{small}~\cite{li2010survey}.
We say an edit distance algorithm is \emph{output-sensitive} if the work is $o(nm)$ when $k=o(n)$.
Many more efficient and/or practical algorithms were proposed in this setting with cost bounds parameterized by $k$~\cite{galil1986improved,galil1987parallel,galil1988data,galil1990improved,harel1984fast,landau1986efficient,landau1988fast,landau1989fast,myers1986nd,myers1986incremental,navarro2001guided}.

Considering the ever-growing data size and plateaued single-processor performance,
it is crucial to consider parallel solutions for edit distance.
Although the problem is simple and well-studied in the sequential setting, we observe a \emph{huge gap} between theory and practice in the parallel setting.
The few implementations we know of~\cite{blelloch2020parlaylib,tchendji2020efficient,yang2010efficient} simply parallelize the $O(nm)$-work sequential algorithm and require $O(n)$ span (longest dependence chain),
which indicates low-parallelism and redundant work when $k\ll n$.
Meanwhile, numerous theoretical parallel algorithms exist~\cite{apostolico1990efficient,babu1997parallel,galil1987parallel,landau1989fast,lu1994parallel,myoupo1999time}, but it remains unknown whether these algorithms are practical (i.e., can be implemented with reasonable engineering effort), and
if so, whether they can yield high performance.
\emph{The goal of this paper is to formally study parallel solutions for edit distance.
By carefully studying existing theoretical solutions, we develop \textbf{new output-sensitive parallel solutions with
good theoretical guarantees and high performance in practice}.
We also conduct in-depth experimental studies on existing and our new algorithms.}

\hide{Edit distance can be solved using dynamic programming (DP), with the states $\dpv[i,j]$ as the edit distance of transforming the first $i$ characters in $A$ to the first $j$ characters in $B$, i.e., the edit distance between $A[1..i]$ and $B[1..j]$. Then $\dpv[i,j]$ can be computed as:
}
The classic dynamic programming (DP) algorithm solves edit distance by using the states $\dpv[i,j]$ as the edit distance of transforming $A[1..i]$ to $B[1..j]$. $\dpv[i,j]$ can be computed as:
\vspace{-.5em}
\begin{align*}
  \dpv[i,j]&=\begin{cases}
  \dpv[i-1,j-1]  \hspace{1.4in} \text{if } A[i]=B[j] \ \text{and} \ i > 0, j > 0 \\
  1+\min(\dpv[i-1,j],\dpv[i-1,j-1],\dpv[i,j-1]) \quad \text{otherwise}
  \end{cases}\\
  \dpv[i,j]&=\max(i,j)\hspace{1.65in} {\text{ if } i=0 \text{ or } j=0}
\end{align*}
A simple parallelization of this computation is to compute all states with the same $i+j$ value in parallel, and process all $i+j$ values in an incremental order~\cite{blelloch2020parlaylib,tchendji2020efficient,yang2010efficient}.
However, this approach has low parallelism as it requires $n+m$ rounds to finish.
Later work~\cite{apostolico1990efficient,babu1997parallel,lu1994parallel} improved parallelism using a \defn{divide-and-conquer (DaC)} approach and achieved $\tilde{O}(n^2)$ work and $\polylog(n)$ span.
These algorithms use the monotonicity of the DP recurrence, and are complicated. 
There are two critical issues in the DaC approaches.
First, to the best of our knowledge, there exist no implementations given the sophistication of these algorithms.
Second, they are not output-sensitive ($\tilde{O}(nm)$ work), which is inefficient when $k\ll n$.

Alternatively,
many existing solutions, 
both sequentially~\cite{galil1986improved,galil1987parallel,galil1988data,galil1990improved,harel1984fast,landau1986efficient,landau1988fast, myers1986nd, myers1986incremental} and in parallel~\cite{galil1987parallel,landau1989fast} use
output-sensitive algorithms, and achieve $\tilde{O}(nk)$ or $\tilde{O}(n+k^2)$ work and $\tilde{O}(k)$ span.
These algorithms view DP table as a grid-like DAG, where each state (cell) $(x,y)$ has three incoming edges from $(x-1,y)$, $(x,y-1)$, and $(x-1,y-1)$ (if they exist).
The edge weight is 0 from $(x-1,y-1)$ to $(x,y)$, when $A[x]=B[y]$, and 1 otherwise.
Then edit distance is equivalent to the shortest path from $(0,0)$ to $(n,m)$.
An example is given in \cref{fig:bfs}.
Since the edge weights can only be 0 or~1, we can use \defn{breadth-first search (BFS)} from the cell $(0,0)$ until $(n,m)$ is reached.
Ukkonen~\cite{ukkonen1985algorithms} further showed that using \defn{longest common prefix (LCP)} queries based on suffix trees or suffix arrays,
the work can be improved to $O(n+k^2)$.
Landau and Vishkin~\cite{landau1989fast} parallelized this algorithm (see~\cref{subsection:algo-BFS}).
While the sequential output-sensitive algorithms have been widely used in practice~\cite{galil1988data,harel1984fast,landau1988fast,myers1986nd,myers1986incremental},
we are unaware of any existing implementations for the parallel version. 

We systematically study parallel output-sensitive edit distance, using both the BFS-based and the DaC-based approaches.
Our first effort is to implement the BFS-based Landau-Vishkin algorithm with our carefully-engineered parallel suffix array (SA) implementation,
referred to as \bfssa.
Although suffix array is theoretically efficient with $O(n)$ construction work, the hidden constant is large.
Thus, we use hashing-based solutions to replace SA for LCP queries to improve the performance in practice. 
We first present a simple approach \bfssimplehash{} in \cref{subsection:algo-hash} that stores a hash value for all prefixes of the input.
This approach has $O(n)$ construction work, $O(\log n)$ per LCP query, and $O(n)$ auxiliary space.
While both \bfssa{} and \bfssimplehash{} take $O(n)$ extra space, such space overhead can be significant in practice---for example, \bfssimplehash{} requires $n$ 64-bit hash values, which is $4\times$ the input size considering characters as inputs, and $32\times$
with even smaller alphabet such as molecule bases (alphabet as $\{A,C,G,T\}$).
To address the space issue, we proposed \bfshash{} using \emph{blocking}.
Our solution takes a user-defined parameter $b$ as the block size, which trades off between space usage and query time.
\bfshash{} limits extra space in $O(n/b)$ by using $O(b\log n)$ LCP query time.
Surprisingly, 
despite a larger LCP cost, our hash-based solutions are consistently faster than \bfssa{} in all real-world test cases, due to cheaper construction.
All of our BFS-based solutions are simple to program.


\begin{table}[t]
    \centering\small
    \vspace{-1em}
    \begin{tabular}{@{}c@{  }@{  }c@{  }@{  }@{  }c@{  }@{  }c@{  }c@{  }@{  }c@{  }@{  }@{  }c@{  }@{  }@{  }c@{}}
      \toprule
      \bf Algorithm & \bf Work & \bf Span  & \bf Space$^*$ &\bf Algorithm & \bf Work & \bf Span  & \bf Space$^*$\\
      \midrule
      \textbf{\bfssa{}} & $O(n+k^2)$ & $\tilde{O}(k)$ & $O(n)$ &\textbf{\bfssimplehash{}$^*$} & $O(n+k^2\log n)$ & $\tilde{O}(k)$ & $O(n)$\\
      \textbf{\dacmm{}} & $O(nk\log k)$ & $\tilde{O}(1)$ & $O(nk)$ &\textbf{\bfshash{}$^*$} & $O(n+k^2b\log n)$ & $\tilde{O}(kb)$ & $O(n/b+k)$\\
      \bottomrule\vspace{-2em}
    \end{tabular}
    \caption{\small \textbf{Algorithms in this paper.} $k$ is the edit distance. $b$ is the block size. $^*$: Monte Carlo algorithms due to the use of hashing. ``Space$^*$'' means auxiliary space used in addition to the input.  Here we assume constant alphabet size for \bfssa{}.
    \label{tab:algo}
    \vspace{-2em}
    }
  \end{table} 

We also study the DaC-based approach and propose a parallel output-sensitive solution.
We propose a non-trivial adaption for the \aalm{} algorithm~\cite{apostolico1990efficient} to make it output-sensitive.
Our algorithm is inspired by the BFS-based approaches, and improves the work from $\tilde{O}(nm)$ to $\tilde{O}(nk)$, with polylogarithmic span.
The technical challenge is that the states in the computation are no longer a rectangle, but an irregular shape (see \cref{fig:bfs,fig:dac_mm}).
We then present a highly non-trivial implementation of this algorithm.
Among many key challenges, we highlight our solution to avoid dynamically allocating arrays in the recursive execution.
While memory allocation is mostly ignored theoretically, in practice it can easily be the performance bottleneck in the parallel setting.
\iffullversion{We refer to this implementation as \dacmm, with details given in \cref{sec:dac-sp,sec:dac-imp,app:aalm-combine-impl}.}
\ifconference{We refer to this implementation as \dacmm, with details given in \cref{sec:dac-sp,sec:dac-imp} and the full version of this paper~\cite{ding2023efficientfull}.}

The bounds of our algorithms (\bfssa{}, \bfssimplehash{}, \bfshash{}, and \dacmm) are presented in \cref{tab:algo}.
We implemented them and show an experimental study in \cref{sec:exp}.
We tested both synthetic and real-world datasets, including DNA,
English text from Wikipedia, and code repositories from GitHub, with string lengths in $10^5$--$10^9$ and varying edit distances,
many of them with real edits (e.g., edit history from Wikipedia and commit history on GitHub).
In most tests, our new \bfshash{} or \bfssimplehash{} performs the best, and their relative performance depends on the value of $k$ and the input patterns.
Our BFS-based algorithms are faster than the existing parallel output-insensitive implementation in ParlayLib~\cite{blelloch2020parlaylib},
even with a reasonably large $k\approx 10^5$.
We believe that our paper is the first systematic study in theory and practice of parallel edit distance,
and we give the first publicly available parallel edit distance implementation
that can process \emph{billion-scale strings} with small edit distance and our code at \cite{edcode}.
\ifconference{Due to page limit, some details are provided in the full version of this paper~\cite{ding2023efficientfull}.}
We summarize our contributions as follows:
\begin{enumerate}[1.]
  \item Two new BFS-based edit distance solutions \bfssimplehash{} and \bfshash{} using hash-based LCP queries.
  Compared to the existing SA-based solution in Landau-Vishkin, our hash-based solutions are simpler and more practical.
  \bfshash{} also allows for tradeoffs between time and auxiliary space.
  \item A new DaC-based edit distance solution \dacmm{} with $O(nk\log k)$ work and polylogarithmic span.
  \item New implementations for four output-sensitive edit distance algorithms: \bfssa{}, \bfssimplehash{}, \bfshash{} and \dacmm{}. Our code is publicly available~\cite{edcode}.
  \item Experimental study of the existing and our new algorithms on different input patterns.
\end{enumerate}


\hide{
Most sequential or parallel solutions are based on computing the DP matrix $\dpv[\cdot,\cdot]$, and the edit distance between $A$ and $B$ is $\dpv[n,m]$.
However, most applications involving minimum edit distance rarely require large edit distances, such as genomic alignment~\cite{delcher2002fast},
git comparison~\cite{spinellis2012git}, and text similarity~\cite{gomaa2013survey}.
Therefore, for large sequence sizes and small minimum edit distance values $k \ll n$,
constructing the complete dynamic programming table is not efficient in terms of time and space complexity.
Thus there are some alternative solutions for the output-sensitive edit distance problems.

The BFS-based algorithms is built on the output-sensitive sequential algorithms, which use the fact that computing the DP matrix is equivalent to
finding the shortest path in a $n\times m$ grid from $(0,0)$ to $(n,m)$.
Each cell $(x,y)$ in the grid has three out-going edges to its left, right, and bottom-right neighbors (if any).
The edge weight is 0 if it is an edge to the bottom-right and $A[x+1]=B[y+1]$ (no edit needed to match $A[x+1]$ and $B[y+1]$),
and 1 otherwise.
As the edges has unit weight (or 0), we can compute the shortest distances by BFS.
If the BFS reaches cell $(x,y)$, the BFS path is a solution to edit $A[1..x]$ to $B[1..y]$, and the length of the path is the edit-distance.
There are two key observations in BFS-based algorithms.
First of all, starting from a cell $(x,y)$, one should go to the bottom-right directly as long as the edge weight is 0.
More specifically, we should always find the longest prefix $s$ starting from $A[x+1]$ and $B[y+1]$ and skip to the cell at $(x+s,y+s)$ with no edit.
Secondly, if the edit distance is $k$, not all cells in the DP matrix need to be computed.
For example, all cells with $|x-y|>k$ are not needed, as going to these cells must require at least $k$ edges of cost 1.
In particular, one can prove that the number of ``meaningful'' cells is $O(k^2)$~\cite{}.
(add a figure to illustrate)
These ideas are used in several sequential algorithms for edit distance and relevant problems~\cite{},
and was also used in some parallel algorithms about approximate string pattern matching~\cite{}.
To parallelize these ideas, the problem boils down to a parallel data structure supporting the ``common-prefix'' query efficiently.
In our paper, we used several parallel data structures.
We first use a parallel suffix-array (SA) based on ~\cite{}, and achieve an edit distance algorithm
(called \bfssa{}) has $O(n+k^2)$ work, $O(k\log k)$ span and $O(n)$ space.
For the hash-based solutions, we first use a simple version \bfssimplehash{} that
has $O(n+k^2)$ work, $O(k\log k)$ span and $O(n\log n)$ space.
Then we optimize the space by compress hash information into blocks, and achieve
\bfshash{}, which has the same work and span as \bfssimplehash{}, but only requires $O(n)$ space.
These algorithms have simple implementations.

For all four algorithms

To achieve practical implementations for edit distance with high parallelism,
we carefully study two commonly-used techniques in existing edit distance algorithms, based on Breadth-First-Search (BFS) and
Divide-and-Conquer (DaC), respectively.
In particular, we study three algorithms: three BFS-based algorithms (\bfssa{}, \bfssimplehash{}, \bfshash{}),
and one DaC-based algorithm (\dacmm). They have different theoretical guarantees in work (total number of operations), span (longest dependence chain of operations) and space (we formally define work and span in \cref{sec:prelim}).

The DaC algorithm (\dacmm{}) is based on a parallel edit-distance algorithm in ~\cite{} with $O(nm\log n)$ work and polylogarithmic span,
and we adapt it to an output-sensitive version that achieves $O(nk\log k)$ work and polylogarithmic span.
The idea of \dacmm{} is also to compute the shortest distance in the grid $G$ (as mentioned above) from $(0,0)$ to $(n,m)$.
We will use a prefix-doubling search to find the value of $k$ (i.e., testing whether $k=1,2,4,8,\dots$ edits is feasible).
For each value of $k$, we only process the stripe of cells in the grid with $|x-y|\le k$, and thus obtain work bound of $\tilde{O}(nk)$.
Specifically, for the DP matrix $D$, we
we compute
the shortest distance for all pairs in $(L\cup U) \times (W\cup R)$,
where $L$ are all cells on the left boundary, $U$ for the upper boundary, $W$ for the lower boundary, and $R$ for the right boundary.
We give an illustration in \cref{fig:dacmm}.
To obtain the shortest distance between such boundary cells in $D$,
the DaC algorithm divides $D$ into $2\times 2$ submatrices, deal with them recursively, and combine
the results on the submatrices.
Importantly, our algorithm in~\cite{} makes use of the monotonicity of the Monge Matrix (as is used in \cite{}) to achieve the work bound.
This algorithm has $O(nk\log k)$ work, polylogarithmic span and $O(k^2P)$ space, where $P$ is the number of processors in the machine. 

We implemented the three algorithms. [Experimental results.]

\xiaojun{We can mention \dacmm{} also works for the weighted case while the other two do not.}
}

\section{Preliminaries}
\label{sec:prelim}

We use $O(f(n))$ \emph{with high probability} (\whp{}) (in $n$) to mean $O(cf(n))$ with probability at least $1-n^{-c}$ for $c \geq 1$.
We use $\tilde{O}(f(n))$ to denote $O(f(n)\cdot \text{polylog}(n))$.
For a string $A$, we use $A[i]$ as the $i$-th character in $A$.
We use \emph{string} and \emph{sequence} interchangeably.
We use $A[i..j]$ to denote the $i$-th to the $j$-th characters in $A$, and $A[i..j)$ the $i$-th to the $(j-1)$-th characters in $A$.
Throughout the paper, we use ``auxiliary space'' to mean space used in addition to the input.

\vspace{-10px}
\subparagraph*{String Edit Distance.}
    Given two strings $A[1..n]$ and $B[1..m]$,
    Levenshtein's Edit Distance~\cite{levenshtein1966binary} between $A$ and~$B$
    is the minimum number of operations needed
     to convert $A$ to $B$ by using insertions, deletions, and substitutions.
     We also call the operations \defn{edits}.
     In this paper, we use \defn{edit distance} to refer to Levenshtein's Edit Distance.
    The classic dynamic programming (DP) algorithm for edit distance
    uses DP recurrence shown in \cref{sec:intro} with $O(mn)$ work and space.

\vspace{-10px}
\subparagraph*{Hash Functions.}\label{prelim:hash-function}
For the simplicity of algorithm descriptions, we assume a perfect hash function for string comparisons,
 i.e., a function $h: S \rightarrow [1, O(|S|)]$ such that $h(x) = h(y) \Longleftrightarrow x = y$.
For any alphabet~$\Sigma$ with size $|\alpha|$, we use a hash function $h(A[l..r]) = \sum_{i=l}^{r}A[i] \times p^{r-i}$ for some prime numbers $p > |\alpha|$, which returns
a unique hash value of the substring $A[l..r]$.
The hash values of two consecutive substrings $S_1$ and $S_2$ can be concatenated as $h([S_1,S_2])=h(S_1)\cdot p^{|S_2|}+h(S_2)$,
and the inverse can also be computed as $h(S_2) = h([S_1, S_2]) - p^{|S_2|} \cdot h(S_1)$.
For simplicity, we denote concatenation and its inverse operation as $\oplus$ and $\ominus$, respectively, as $h([S_1,S_2])=h(S_1)\oplus h(S_2)$ and $h(S_2)=h([S_1, S_2])\ominus h(S_1)$.
We assume perfect hashing for theoretical analysis.
In practice, we use $p$ as a large prime and modular arithmetic to keep the word-size hash values.
In our experiment, we compare different approaches and validate that our implementations are correct in all test cases.
However, collisions are possible for other datasets, since different strings may be mapped to the same hash value. 
If such cases arise, one can either use multiple hash functions for a better success rate in practice, or use the idea of Hirschberg's algorithm~\cite{hirschberg1975linear} to generate the edit sequence and run a correctness check (and restart with another hash function if failed). 

\vspace{-10px}
\subparagraph*{Longest Common Prefix (LCP).}
For two sequences $\seqA[1..n]$ and $\seqB[1..m]$, the Longest Common Prefix (LCP) query at position $x$ in $A[1..n]$ and position $y$ in $B[1..m]$
is the longest substring starting from $A[x]$ that match a prefix starting from $B[y]$.
With clear context,
we also use the term ``LCP'' to refer to the length of the LCP,
i.e., $\mb{LCP}(\seqA, \seqB, x, y)$ is the length of the longest common prefix substring starting from $A[x]$ and $B[y]$ for $A$ and $B$.

\vspace{-10px}
\subparagraph*{Computational Model.}
We use the \defn{work-span model} in the classic multithreaded model with \defn{binary-forking}~\cite{arora2001thread,blelloch2020optimal,BL98}. 
We assume a set of \thread{}s that share the memory.
Each \thread{} acts like a sequential RAM plus a \forkins{} instruction
that forks two child threads running in parallel.
When both child threads finish, the parent thread continues. 
A parallel-for is simulated by \forkins{} for a logarithmic number of steps.
A computation can be viewed as a DAG (directed acyclic graph).
The \defn{work $\boldsymbol{W}$} of a parallel algorithm is the total number of operations,
and the \defn{span (depth) $\boldsymbol{S}$} is the longest path in the DAG.
The randomized work-stealing scheduler can execute such a computation in $W/P+O(S)$ time \whp{} in $W$ on $P$ processors~\cite{arora2001thread,BL98,gu2022analysis}.
\vspace{-10px}
\subparagraph*{Suffix Array.}
The suffix array (SA) \cite{manber1993suffix} is a lexicographically sorted array of the suffixes of a
string, usually used together with the longest common prefix (LCP) array,
which stores the length of LCP between every adjacent pair of suffixes.
The SA and LCP array can be built in parallel in $O(n)$ work and $O(\log^2n)$ span \whp{}~\cite{karkkainen2003simple,shun2014fast}.

In edit distance, we need the LCP query between $A[x..n]$ and $B[y..m]$ for any $x$ and $y$.
This can be computed by building the SA and LCP arrays for a new string $C[1..n+m]$ that concatenates $A[1..n]$ and $B[1..m]$.
The LCP between any pair of suffixes in $C$ can be computed by a range minimum query (RMQ) on the LCP array,
which can be built in $O(n+m)$ work and $O(\log(n+m))$ span~\cite{blelloch2020optimal}.
Combining all pieces gives the following theorem:

\begin{lemma}\label{thm:parallelsa}
  Given two strings $A[1..n]$ and $B[1..m]$, using a suffix array,
  the longest common prefix (LCP) between any two substrings
  $A[x..n]$ and $B[y..m]$ can be reported in $O(1)$ work and span,
  with $O(n+m)$ preprocessing work and $O(\log^2(n+m))$ span \whp{}.
\end{lemma}

\section{BFS-based Algorithms}
\label{subsection:algo-BFS}

\subsection{Overview of Existing Sequential and Parallel BFS-based Algorithms}\label{sec:algo-BFS-prior}
Many existing output-sensitive algorithms~\cite{galil1986improved,galil1987parallel,galil1988data, galil1990improved,harel1984fast,landau1986efficient,landau1988fast,landau1989fast,myers1986nd, myers1986incremental} are based on breadth-first search (BFS).
These algorithms view the DP matrix for edit distance as a DAG, as shown in \cref{fig:bfs}.
In this section, we use $x$ and~$y$ to denote the row and column ids of the cells in the DP matrix, respectively.
Each state (cell) $(x,y)$ has three incoming edges from $(x-1,y)$, $(x,y-1)$, and $(x-1,y-1)$ (if they exist).
The edge weight is 0 from $(x-1,y-1)$ to $(x,y)$ when $A[x]=B[y]$, and 1 otherwise.
Then edit distance is equivalent to the \shortestdis{} from $(0,0)$ to $(n,m)$.

Since the edge weights are 0 or 1, we can use a special breadth-first search (BFS) to compute the \shortestdis{}.
In round $t$, we process states with edit distance $t$.
The algorithm terminates when we reach cell $(n,m)$.
First observed by Ukkonen~\cite{ukkonen1985algorithms}, in the BFS-based approach, not all states need to be visited.
For example, all states with $|x-y|>k$ will not be reached before we reach $(n,m)$ with edit distance $k$, since they require more than $k$ edits.
Thus, this BFS will touch at most $O(kn)$ cells, leading to $O(kn)$ work.

Another key observation is that starting from any cell $(x,y)$,
if there are diagonal edges with weight 0, we should always follow the edges until a unit-weight edge is encountered.
Namely, we should always find the longest common prefix (LCP) from $A[x+1]$ and $B[y+1]$,
and skip to the cell at $(x+p,y+p)$ with no edit, where $p$ is the LCP length.
This idea is used in Landau and Vishkin~\cite{landau1989fast} on parallel approximate string matching, and
we adapt this idea to edit distance here. 
Using the modified parallel BFS algorithm by Landau-Vishkin~\cite{landau1989fast} (shown in \cref{algo:bfs_in_code}), only $O(k^2)$ states need to be processed---on each diagonal and for each edit distance $t$,
only the last cell with $t$ edits needs to be processed (see \cref{fig:bfs}). 
Hence, the BFS runs for $k$ rounds on $2k+1$ diagonals, which gives the $O(k^2)$ bound above.
In the BFS algorithm, we can label each diagonal by the value of $x-y$.
In round $t$, the BFS visits a \emph{frontier} of cells $f_t[\cdot]$, where $f_t[i]$ is the cell with edit distance $t$ on diagonal $i$, for $-t\le i\le t$.
We present the algorithm in \cref{algo:bfs_in_code} and an illustration in \cref{fig:bfs}.
Note that in the implementation, we only need to maintain two frontiers (the previous and the current one), which requires $O(k)$ space.
We provide more details about this algorithm \iffullversion{in \cref{app:lv-algo}.}\ifconference{in the full version~\cite{ding2023efficientfull}.}
If the LCP query is supported by suffix arrays, we can achieve $O(n+k^2)$ work and $O(\log n+k \log k)$ span for the edit distance algorithm.


\begin{figure}
  \hspace{-.3in}\begin{minipage}[H]{0.5\textwidth}
  \vspace{-.1in}
    \begin{flushleft}
    \begin{figure}[H]
      \includegraphics[page=1,width=1.05\columnwidth]{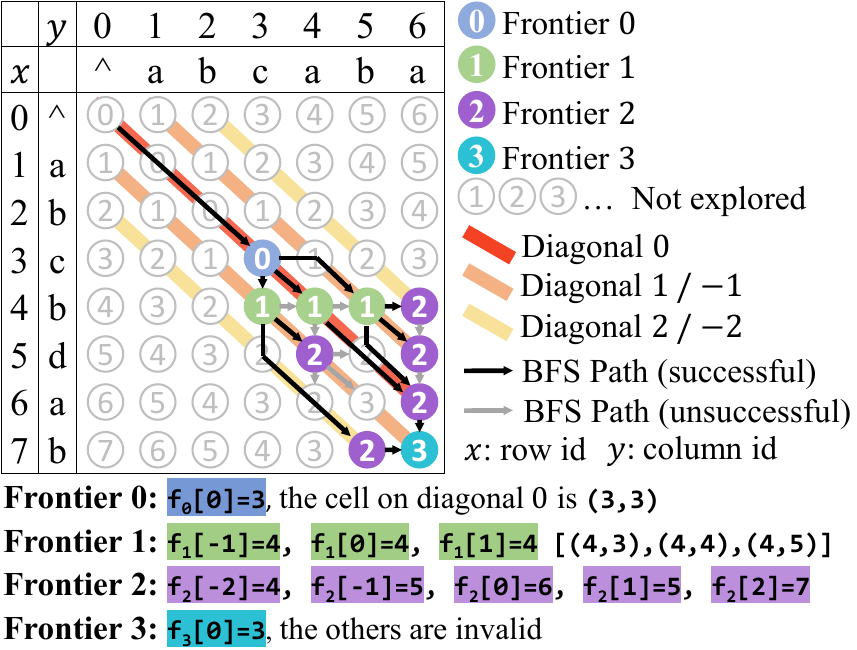}
      \caption{\small \textbf{BFS-based edit distance on $A[1..n]$ and $B[1..m]$.} A more detailed description is in 
      \iffullversion{\cref{app:lv-algo}.}
      \ifconference{the full version~\cite{ding2023efficientfull}.}
      $f_t[i]$ is the row-id of the last cell on diagonal $i$ with edit distance $t$ (frontier $t$), representing
      cell $(f_t[i], f_t[i]-i)$.
      \label{fig:bfs}}
    \end{figure}
    \end{flushleft}
\end{minipage}%
\hfill\begin{minipage}[H]{0.5\textwidth}
\vspace{-3em}
    \begin{flushright}
      \begin{algorithm}[H]
      \SetInd{0.5em}{0.5em}
        \fontsize{9pt}{9pt}\selectfont
\caption{\small BFS-based parallel edit distance~\cite{landau1989fast} \label{algo:bfs_in_code}}

  \DontPrintSemicolon
  $f_0[0]\gets$\longestprefix$(A[1..n],B[1..m])$\tcp*[f]{Starting point}\label{line:bfs-starting}\\
  $t \gets 0$\\
  \While {$f_t[n-m]\ne n$\label{line:bfs-outer-loop}} {
    $t\gets t+1$\\
    \tcp{Find new frontier for diagonal $i$}
    \parForEach{$-t\le i \le t$}{
      $f_{t}[i]\gets f_{t-1}[i]$ \tcp*[f]{Start from the last cell}\\
      \ForEach {$\langle\dx,\dy\rangle\in\{\langle0,1\rangle,\langle1,0\rangle,\langle1,1\rangle\}$} {
        \tcp{The previous cell is from diagonal $j$ }
        \tcp{$j=(x-dx)-(y-dy)=i-dx+dy$}
        $j\gets i-dx+dy$\label{line:bfs:j}\\
        \If{$|j|\le t-1$} {
          The row id $x\gets f_{t-1}[j]+dx$\label{line:bfs:x}\\
          The column id $y\gets x-i$\label{line:bfs:y}\\
          \tcp{Skip the common prefix}
          $x\gets x+$\longestprefix$(A[x+1..n],B[y+1..m])$\label{line:bfs:addp}\\

          \tcp{Keep the largest row id}
          $f_{t}[i]\gets\max(f_{t}[i],x)$\\
        }
      }
    }
  }
  \Return{$t$}\\
      \end{algorithm}
    \end{flushright}
\end{minipage}
\vspace{-3.8em}
\end{figure}

\hide{
As we described in \cref{sec:intro}, the search process will be bounded in the range around the
diagonal direction with width $O(k)$. Therefore, a frontier with size $2k$ is maintained for the search range of next round.
In each search round, we find the cells in the frontier in parallel, and continue the following BFS precess.
Firstly, the position in the DP matrix with the id of row and col can be located.
Then we query the \LCP () by enumerating the three directions (), corresponding the table updating according to the recurrence function.
Getting the \LCP value, we compare it with the current position value and keep the largest row ID (),
that is the farthest distance in the diagonal direction of the DP matrix shown in \cref{fig:bfs}(b).
Finally we update the frontier with the values in $g[i]$.
}

\vspace{-10px}
\subparagraph*{Algorithm Based on Suffix Array (\bfssa{}).}\label{para:algo-sa}
Using the SA algorithm in~\cite{karkkainen2003simple} and the LCP algorithm in \cite{shun2014fast} for Landau-Vishkin gives the claimed bounds in \cref{tab:algo}.
We present details about our SA implementation in \cref{sec:impl-sa}.

\hide{
\begin{theorem}
  The \bfssa{} algorithm computes the edit distance between two sequences of length $n$ and $m\le n$ in $O(n+k^2)$ work and $\tilde{O}(k)$ span,
  where $k$ is the output size (fewest possible edits).
\end{theorem}
}

\subsection{Algorithm Based on String Hashing (\bfssimplehash{})}\label{subsection:algo-hash}

Although \bfssa{} is theoretically efficient with $O(n)$ preprocessing work to construct the SA, the hidden constant is large.
For better performance, we consider string hashing as an alternative for SA.
Similar attempts (e.g., locality-sensitive hashing) have also been used in approximate pattern matching problems~\cite{marccais2019locality,mccauley:LIPIcs.ICDT.2021.21}.
In our pursuit of exact output-sensitive edit distance computation, we draw inspiration from established string hashing algorithms, such as the Rabin-Karp algorithm (also known as rolling hashing)~\cite{karp1987efficient}.
We will first present a simple hash-based solution \bfssimplehash{} with $O(n)$ preprocessing cost and $O(n)$ auxiliary space.
Then later in \cref{subsec:block-hashing}, we will present \bfshash{}, which saves auxiliary space by trading off more work in LCP queries.

As mentioned in \cref{prelim:hash-function},
the hash function $h(\cdot)$ maps any substring $A[l..r]$ to a unique hash value,
which provides a fingerprint for this substring in the LCP query. 
The high-level idea is to binary search the query length, using the hash value as validation.
We precompute the hash values for all prefixes, i.e., $T_A[x]=h(A[1..x])$ for the prefix substring $A[1..x]$ (similar for $B$).
They can be computed in parallel by using any scan (prefix-sum) operation~\cite{Blelloch89} with $O(n)$ work and $O(\log n)$ span. 
We can compute $h(A[l..r])$ by $T_A[r]\ominus T_A[l-1]$.

With the preprocessed hash values, we dual binary search the LCP of $A[x..n]$ and $B[y..m]$.
We compare the hash values starting from $A[x]$ and $B[y]$ with
chunk sizes of $1,2,4,8,\dots$, until we find value $l$, such that $A[x..x+2^{l})=B[y..y+2^{l})$,
but $A[x..x+2^{l+1})\ne B[y..y+2^{l+1})$. By doing this with $O(\log n)$ work,
 we know that the LCP of $A[x..n]$ and $B[y..m]$ must have a length in the range $[2^l,2^{l+1})$.
 We then perform a regular binary search in this range, which costs another $O(\log n)$ work. This indicates
 $O(\log n)$ work in total per LCP query. Combining the preprocessing and query costs, we present the cost bounds
 of \bfssimplehash{}:

\begin{theorem}
  \label{thm:bfs-simple-hash}
  \bfssimplehash{} computes the edit distance between two sequences of length $n$ and $m\le n$ in $O(n+k^2\log n)$ work, $\tilde{O}(k)$ span, and $O(n)$ auxiliary space,
  where $k$ is the output size (fewest possible edits).
\end{theorem}
\bfssimplehash{} is simple and easy to implement. Our experimental results indicate that its simplicity also allows for a reasonably good performance in practice for most real-world input instances.
However, this algorithm uses $n$ 64-bit integers as hash values, and such space overhead may be a concern in practice.
This is more pronounced when the input is large and/or the alphabet is small (particularly when each input element can be represented with smaller than byte size),
as the auxiliary space can be much larger than the input size.
This concern also holds for \bfssa{} as several $O(n)$-size arrays are needed during SA construction.
Note that for shared-memory parallel algorithms, space consumption is also a \emph{key constraint}---if an algorithm is slow, we can wait for longer;
but if data (and auxiliary data) do not fit into the memory, then this algorithm is not applicable to large input at all.
In this case, the problem size that is solvable by the algorithm is limited by the space overhead, which makes the improvement from parallelism much narrower.
Below we will discuss how to make our edit distance algorithms more space efficient.

\subsection{Algorithm Based on Blocked-Hashing (\bfshash{})}\label{subsec:block-hashing}

In this section, we introduce our \bfshash{} algorithm that provides a more space-efficient solution by trading off worst-case time (work and span).
Interestingly, we observed that on many data sets, \bfshash{} can even outperform \bfssimplehash{} and other opponents due to faster construction time,
and we will analyze that in \cref{sec:exp}.

To achieve better space usage, we divide the strings into blocks of size $b$.
As such, we only need to store the hash values for prefixes of the entire blocks $h(A[1..b]), h(A[1..2b]), \cdots, \\h(A[1..\lfloor (n / b) \rfloor \cdot b])$.
Our idea of blocking is inspired by many string algorithms  (e.g.,~\cite{bender2000lca}).
Using this approach, we only need auxiliary space to store $O(n/b)$ hash values, and thus we can control the space usage using the parameter $b$.
To compute these hash values, we will first compute the hash value for each block, and run a parallel scan (prefix sum on $\oplus$) on the hash values for all the blocks.
Similar to the above, we refer to these arrays as $T_A[i]=h(A[1..ib])$ (and $T_B[i]$ accordingly), and call them \defn{\hashtabfull{s}}.

\begin{figure*}[!t]
    \centering
    \vspace{-1.0em}
    \includegraphics[width=1.0\columnwidth,page=2]{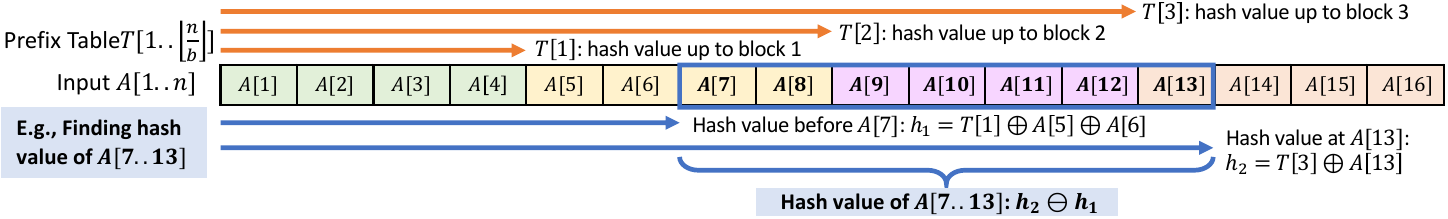}
    \vspace{-2.0em}
    \caption{\small \textbf{The illustrations of \hashtab{} values and one specific query}, with key concepts shown when computing the hash value of a range using a \hashtab. \label{fig:hash-illustration}
    \vspace{-0.5em}
    }
  \end{figure*} 

\hide{
\begin{algorithm}[t]
\fontsize{8pt}{9pt}\selectfont
\caption{\small The \hashtab{}\label{algo:hashtab}}

\SetKw{Break}{break}
\SetKw{Kwand}{and}
\SetKw{Kwor}{or}

\KwIn{Two sequences $A_{1..n}$ and $B_{1..m}$.}
\SetKwInOut{Global}{Global Variables}
    \vspace{0.5em}


\DontPrintSemicolon
\Global{
\block{} size: $b= O(\log n)$\\
Number of \block{s} in $A$: $k_A=n/b$\\
Number of \block{s} in $B$: $k_B=m/b$\\
}

\MyFunc{\textsc{Construct}$(A,B)$} {
  $T_A[\cdot][\cdot]\gets$\textsc{Build}$(A)$\\
  $T_B[\cdot][\cdot]\gets$\textsc{Build}$(B)$
}

\MyFunc{\upshape\textsc{Build}$(A)$} {
  $k\gets |A|/b$\\
  \parForEach{$j\gets 1$ to $k$} {
    $T[1][j]\gets \hashfunc(A[j..j+b])$
  }
  \For{$i\gets 1$ to $\log k$} {
    \parForEach{$j\gets 1$ to $k$} {
      $T[i][j]\gets T[i-1][j] \hashcombine T[i-1][j+2^{i-1}]$
    }
  }
  \Return {$T[\cdot][\cdot]$}
}

\tcp{Longest Prefix starting from $A[x]$ and $B[y]$}
\MyFunc{\upshape\longestprefix$(x,y)$} {
 $l\gets 1$\\
 \While{$((l+x)\mmod b)\ne 0$ or $((l+y)\mmod b)\ne 0$}{
 \lIf{$A[l+x]\ne B[l+y]$}{
 \Return{$(l-1)$}
 } \lElse {
   $l\gets l+1$
 }
 }
 \tcp{The current pointer in $A$ $(l+x)$ is $\remain{A}$ away from the next \block{};
 the next \block{} is the $\nextblock{A}$-th \block{}}
 $\nextblock{A}=\lceil(l+x)/b\rceil$\\
 $\remain{A}=\nextblock{A}\cdot b-(l+x)$\\
 \tcp{Compute $\remain{B}$ and $\nextblock{B}$ for $B$ similarly}
 $\nextblock{B}=\lceil(l+y)/b\rceil$\\
 $\remain{B}=S_B\cdot b-(l+y)$\\
 $J\gets 0$\\
 \While{$J<\min(\log k_A,\log k_B)$} {
   \tcp{$\suffixhash{A}[l+x]$: hash value from $l+x$ to the end of this \block{}. }
   \tcp{$T[J][\nextblock{A}]$: hash value of the next $2^J$ \block{s}. }
   \tcp{$\prefixhash{A}[\nextblock{A}\cdot b+\remain{B}]$: hash value of the first $\remain{B}$ elements in the next \block{} (the $(\nextblock{A}+1)$-th). This is to align with the subsequence in $B$.}
   $x\gets \suffixhash{A}[l+x] \hashcombine T_A[J][\nextblock{A}] \hashcombine \prefixhash{A}[\nextblock{A}\cdot b+\remain{B}]$\\
   $y\gets \suffixhash{B}[l+y] \hashcombine T_B[J][\nextblock{B}] \hashcombine \prefixhash{B}[\nextblock{B}\cdot b+\remain{A}]$\\
   \lIf{$x\ne y$} { \Break{}
   }
   $J\gets J+1$
 }
 \tcp{Found the current common prefix of length $t$}
 $t\gets b\times 2^{J-1}+\remain{A}+\remain{B}$\\
 \Return{$t$ + \longestprefix$(x+t, y+t)$}
}
\MyFunc{\upshape{Compare}$(x, y, J)$} {

}

\end{algorithm}
}

\newcommand{\firstjump}{{{l_1}}}
\newcommand{\secondjump}{{{l_2}}}


\begin{algorithm}[t]
\fontsize{9pt}{9pt}
\selectfont
\SetInd{0.6em}{0.6em}

\caption{{\small The \hashtab{} for finding the longest common prefix of $A[1..n]$ and $B[1..m]$\label{algo:hashtab}}}

\SetKw{Break}{break}
\SetKw{Kwand}{and}
\SetKw{Kwor}{or}
\SetNlSkip{0.2em}

\SetKwInOut{note}{Notes}


\DontPrintSemicolon
\begin{minipage}[t]{.42\columnwidth}
\fontsize{9pt}{9pt}
\selectfont
\tcp{Table construction}
\MyFunc{\fname{Construct}$(A,B)$} {
  $T_A[\cdot]\gets$\fname{Build}$(A)$\;
  $T_B[\cdot]\gets$\fname{Build}$(B)$\;
}
\codeskip

\tcp{The \hashtab{} building process}
\MyFunc{\upshape\fname{Build}$(A)$} {
  $w\gets \lfloor |A|/\bsize{} \rfloor$\;
  $ T[0] \gets 0$ \;
  \parForEach{$j\gets 1$ to $w$} {
    $T[j]\gets \hashfunc(A[(j - 1)\bsize+1\,..\,j\bsize{}])$
  }
  \fname{Scan}$(T)$ \;
  \Return {$T[\cdot]$}
}
\codeskip

\tcp{Get hash value for prefix sub-\\sequence $A[1..x]$}
\MyFunc{\fname{GetHash}$(A, T_A, x)$} {
  \lIf{$x = 0$}{\Return{$0$}}
  $r \gets \lfloor (x - 1) / b + 1\rfloor$ \;
  $\bar{h} \gets T_A[r]$\;
  \For {$i \gets r \cdot b+1$ to $x$} {
    $\bar{h} \gets \bar{h} \hashcombine h(A[i])$ \label{line:lastcheck}
  }
  \Return {$\bar{h}$}
}
\codeskip
\end{minipage}\hfill\begin{minipage}[t]{0.55\columnwidth}
  \fontsize{9pt}{9pt}
  \selectfont

\tcp{Compare the subsequences $A[x..x+l]$ and $B[y..y+l]$}
\MyFunc{\fname{Compare}$(A, B, x, y, l)$} {
    $h_A \gets \fname{GetHash}(A, T_A, x+l) \hashdiff \fname{GetHash}(A, T_A, x-1)$

    $h_B \gets \fname{GetHash}(B, T_B, y+l) \hashdiff \fname{GetHash}(B, T_B, y-1)$

  \Return{$h_A = h_B$}
}

\codeskip
\codeskip
\tcp{Longest Common Prefix from $A[x]$ and $B[y]$}
\MyFunc{\upshape\longestprefix$(A, B, x,y)$} {
 $\firstjump\gets 0$\;
 \tcp{Find $\firstjump$, s.t. the LCP is between $2^{\firstjump}$ to $2^{\firstjump+1}$ \block{s}}
 \While{$x + 2^{\firstjump} < n$ \Kwand{} $y + 2^{\firstjump} <m$\label{line:whilestart}} {
   \lIf{\fname{Compare}$(A,B,x,y,\firstjump) = \false$} { \Break{}
   }
   $\firstjump\gets \firstjump+1$\label{line:whileend}
 }
 \tcp{Trivial binary search process on the range $[2^{\firstjump}, 2^{\firstjump + 1})$}



  $s \gets 2^{\firstjump}, t \gets 2^{\firstjump + 1}$ \;
  \While{$s < t$}{
    \If{\fname{Compare}$(A, B, x, y, \lfloor (s + t) / 2 \rfloor) = \false$}{
      $t \gets  \lfloor (s + t) / 2 \rfloor$
    } \lElse {
      $s \gets \lfloor (s + t) / 2 \rfloor + 1$
    }
  }
  \Return{$s$}

}
\end{minipage}
\end{algorithm} 
We now discuss how to run LCP with only partial hash values available.
The \longestprefix{} function in \cref{algo:hashtab} presents the process to find the \LCP{} of $A[x..n]$ and $B[y..m]$ using the \hashtab{s}.
We present an illustration in \cref{fig:hash-illustration}.
We will use the same dual binary search approach to find the LCP of two strings.
Since we do not store the hash values for all prefixes,
we use a function \fname{GetHash}$(A, T_A, x)$ to compute $h(A[1..x])$.
\hide{To compare the hash values starting from arbitrary positions in $A$ and $B$,
in \cref{algo:hashtab}, we use a simple helper function \fname{GetHash}$(A, T_A, x)$ that computes $h(A[1..x])$ using $T_A$.}
We can locate the closest precomputed hash value and use $r$ as the previous block id before $x$.
Then the hash value up to block $r$ is simply $\bar{h}=T_A[r]$.
We then concatenate the rest characters to the hash value (i.e., return $\bar{h}\oplus h(A[rb+1])\oplus\cdots \oplus h(A[x])$).
In this way, we can compute the hash value of any prefixes for both $A$ and $B$, and plug this scheme into the dual
binary search in \bfssimplehash{}. In each step of dual binary search, the concatenation of hash value can have at most $b$ steps,
and thus leads to a factor of $b$ overhead in query time than \bfssimplehash{}.

 \begin{theorem}
   \bfshash{} computes the edit distance between two sequences of length $n$ and $m\le n$ in $O(n+k^2\cdot b\log n)$ work and $\tilde{O}(kb)$ span, using $O(n/b+k)$ auxiliary space, where $k$ is the output size (fewest possible edits).
 \end{theorem}

The term $k$ in space usage is from the BFS (each frontier is at most size $O(k)$).
$O(b\log n)$ is the work for each LCP query.
Note that this is an upper bound---if the LCP length $L$ is small, the cost can be significantly smaller (a tighter bound is $O(\min(L,b\log L))$).
\cref{sec:exp} will show that for normal input strings where the LCP lengths are small in most queries,
the performance of \bfshash{} is indeed the fastest, although for certain input instances when the worst case is reached, the performance is not as good.

\section{The Divide-and-Conquer Algorithms}
\label{sec:dac-sp}

Our parallel output-sensitive algorithm \dacmm{} is inspired by the \aalm{} algorithm~\cite{apostolico1990efficient}, and also uses it as a subroutine.
We first overview the \aalm{} algorithm, and introduce our algorithm in details.
We assume $m=n$ is a power of 2 in this section for simple descriptions, but both our algorithm and \aalm{} work for any $n$ and $m$.

\vspace{-10px}
\subparagraph*{The \aalm{} Algorithm.}
As described above, the edit distance problem can be considered as a \shortestdis \space (\SP{}) problem from
the top-left cell $(0,0)$ to the bottom-right cell $(n,n)$ in the DP matrix $\dpv$.
Instead of directly computing the \SP{} from $(0,0)$ to $(n,n)$, \aalm{}
computes pairwise \SP{} between any cell on the left/top boundaries and the bottom/right boundaries (i.e., those on $L\cup U$ to $W\cup R$ in \cref{fig:dac_mm}(a)).
We relabel all cells in $L\cup U$ as a sequence $v=\{v_0, v_1, \dots v_{2n}\}$ (resp., $W\cup R$ as $u=\{u_0, u_1 \cdots, u_{2n}\}$), as shown in \cref{fig:dac_mm}.
Therefore, for the DP matrix $G$, the pairwise \SP{} between $v$ and $u$ forms a $(2n+1) \times (2n+1)$ matrix.
We call it the \defn{\SP{} matrix} of $G$, and denote it as $\dist_\dpv$.
\aalm{} uses a divide-and-conquer approach.
It first partitions $\dpv$ into four equal submatrices $\dpv_1$, $\dpv_2$, $\dpv_3$, and $\dpv_4$ (See \cref{fig:dac_mm}(b)),
and recursively computes the \SP{} matrices for all $\dpv_i$.
We use $\dist_i$ to denote the \SP{} matrix for $\dpv_i$.
In the ``conquer'' step, the \aalm{} algorithm uses a \emph{Combine} subroutine to combine two \SP{} matrices into one if they share a common boundary (our algorithm
also uses this subroutine).
For example, consider combining $\dpv_1$ and $\dpv_2$.
We still use $v_i$ and $u_j$ to denote the cells on the left/top and bottom/right boundaries of $\left(\begin{matrix}
                                                                                                         \dpv_1 \\
                                                                                                         \dpv_2
                                                                                                       \end{matrix}\right)$
(see \cref{fig:dac_mm}(c)), and denote the cells on the common boundary of $\dpv_1$ and $\dpv_2$ as $w_1,\cdots,w_{n/2}$, ordered from left to right.
For any pair $v_i$ and $u_j$, if they are in the same submatrix, we can directly get the \SP{} from the corresponding \SP{} matrix.
Otherwise, WLOG assume $v_i\in \dpv_1$ and $u_j\in \dpv_2$, then we compute the \SP{} between them by finding $\min_{l} \dist_1[i,l]+\dist_2[l,j]$,
i.e., for all $w_l$ on the common boundary, we attempt to use the \SP{} between $v_i$ to $w_l$, and $w_l$ to $u_j$, and find the minimum one.
Similarly, we can combine $\dist_3$ with $\dist_4$, and $\dist_{1\cup 2}$ with $\dist_{3\cup 4}$, and eventually get $\dist_\dpv$.
We note that the \fname{Combine} algorithm, even theoretically, is highly involved.
At a high level, it uses the Monge property of the \shortestdis{} (the monotonicity of the DP recurrence), and we refer the readers to~\cite{apostolico1990efficient} for a detailed algorithm description and theoretical analysis.
In \cref{sec:dac-imp}, we highlight a few challenges and our solutions for implementing this highly complicated algorithm.
Theoretically, combining two $n\times n$ \SP{} matrices can be performed in $O(n^2)$ work and $O(\log^2 n)$ span, which gives
$O(n^2\log n)$ work and $O(\log^3 n)$ span for \aalm{}. 

\begin{figure*}[t]
    \centering
    \vspace{-1.5em}
    \includegraphics[width=.9\columnwidth]{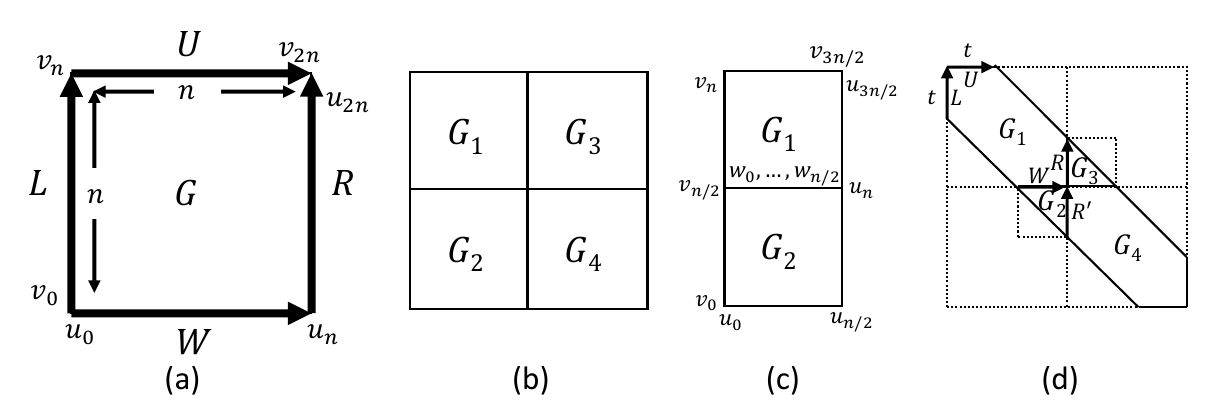}
    \vspace{-1em}
    \caption{\small \textbf{The illustrations of the key concepts and notation in the \aalm{} algorithm described in~\cref{sec:dac-sp}.}
    \label{fig:dac_mm}\vspace{-.5em}}
  \end{figure*}

\vspace{-10px}
\subparagraph*{Our algorithm.}
The \aalm{} algorithm has $\tilde{O}(n^2)$ work ($\tilde{O}(nm)$ if $n\ne m$) and polylogarithmic span, which
is inefficient in the output-sensitive setting. 
As mentioned in \cref{sec:algo-BFS-prior}, only a narrow width-$O(k)$ diagonal area in $G$ is useful (\cref{fig:dac_mm}(d)).
We thus propose an output-sensitive \dacmm{} algorithm adapted from the \aalm{} algorithm.
We follow the same steps in \aalm{},
but restrict the paths to the diagonal area, although the exact size is unknown ahead of time.
We first present the algorithm to compute the \shortestdis{} on the diagonal region with width $2t+1$ as function \fname{Check}$(t)$ in \cref{algo:dac-mm},
which restricts the search in diagonals $-t$ to $t$.
First, we divide such a region into four sub-regions (see \cref{fig:dac_mm}(d)).
Two of them ($\dpv_1$ and $\dpv_4$) are of the same shape, and the other two of them ($\dpv_2$ and $\dpv_3$) are triangles.
For $\dpv_2$ and $\dpv_3$, we use the \aalm{} algorithm to compute their \SP{} matrices by aligning them to squares.
For $\dpv_1$ and $\dpv_4$, we process them recursively,
until the base case where the edge length of the matrix is smaller than $t$ and they degenerate to squares,
in which case we apply the \aalm{} algorithm.
Note that even though the width-$(2t+1)$ diagonal stripe is not a square ($\dpv_1$ and $\dpv_4$ are also of the same shape),
the useful boundaries are still the left/top and bottom/right boundaries ($L\cup U$ and $W\cup R$ in \cref{fig:dac_mm}(d)).
Therefore, we can use the same \fname{Combine} algorithm as in \aalm{} to combine the \SP{} matrices.
For example, in \cref{fig:dac_mm}(d), when combining $G_1$ with $G_2$, we obtain the pairwise distance between $L\cup U$ and
$R\cup R'$ using the common boundary $W$. We can similarly combine all $G_1,G_2,G_3,$ and $G_4$ to get the \SP{} matrix for $G$.

\hide{
\begin{algorithm}[t]
\fontsize{9pt}{9pt}
\selectfont

\caption{{\small Divide-and-Conquer based algorithm \label{algo:dac-mm}}}

\SetKw{Break}{break}
\SetKw{Kwand}{and}
\SetKw{Kwor}{or}
\SetKw{And}{and}
\SetKw{Or}{or}
\SetNlSkip{0.2em}

\SetKwInOut{note}{Notes}


\DontPrintSemicolon
\begin{minipage}[t]{.47\columnwidth}
\fontsize{9pt}{9pt}
\selectfont
\tcp{Find the edit distance between $A$ and $B$}
\MyFunc{\upshape \mf{\dacmm{}}$(A[1..n], B[1..m])$} {
$t \gets \max(1, n-m)$ \;
\While{\true} {
    $\dist \gets \mf{\dacmm{}}(0, n, 0, m, t, t)$ \;
    \lIf {$\dist[t][t] \leq t$} {
        \Break    }
    $t = \min(2\times t, n)$ \;
}
\Return{$\dist[t][t]$}\;
}
\codeskip
\MyFunc{\upshape \mf{\dacmmrec{}}$(i, n, j, m, t_1, t_2)$} {
    \If {$n/2 < t_1$ \Or $m/2 < t_2$} {
        $\dist \gets$ computed by the \aalm{} algorithm \;
        \Return {$\dist$}
    }
    $n_1, n_2 \gets$ divide $n$ by half \;
    $m_1, m_2 \gets$ divide $m$ by half \;
    $\dist_1 \gets \mf{\dacmmrec{}}(i, n_1, j, m_1, t_1, t_2)$ \;
    $\dist_2 \gets$ computed by the \aalm{} algorithm \;
    $\dist_3 \gets$ computed by the \aalm{} algorithm \;
    $\dist_4 \gets \mf{\dacmmrec{}}(i+n_1, n_2, j+m_1, m_2, m_1+t_1-n_1, n_1-m_1+t_2)$ \;
    $\dist_{1\cup 2} \gets \mf{Combine}(\dist_1, \dist_2)$ \;
    $\dist_{3\cup 4} \gets \mf{Combine}(\dist_3, \dist_4)$ \;
    $\dist \gets \mf{Combine}(\dist_{1\cup 2}, \dist_{3\cup 4})$ \;
    \Return{$\dist$}\;
}
\end{minipage}\hfill\begin{minipage}[t]{0.47\columnwidth}
\fontsize{9pt}{9pt}
\selectfont
\tcp{The combine function}
\MyFunc{\upshape \mf{Combine}$(\dist_i, \dist_j)$} {
    $\dist \gets$ Initialize the distance matrix \;
    Copy the pairwise distances that does not cross the common boundary from $\dist_i$ \And $\dist_j$ to $\dist$ \;
    $n \gets$ length of $\dist_i$ \;
    $m \gets$ length of $\dist_j$ \;
    $p \gets$ the largest power of two that is smaller than $n$ \;
    $q \gets$ the largest power of two that is smaller than $m$ \;
    \xiaojun{Add the details of the compute function.} \;
    \While {\true} {
        $\mf{ComputeOddEven}(p, q)$ \;
        $\mf{ComputeEvenOdd}(p, q)$ \;
        $\mf{ComputeOddOdd}(p, q)$ \;
        \lIf {$p=1$ \And $q=1$} {\Break}
        $p = \max(p/2, 1)$ \;
        $q = \max(q/2, 1)$ \;
    }
    \Return{$\dist$}
}
\end{minipage}

\end{algorithm}
}

\begin{algorithm}[t]
\fontsize{9pt}{9pt}
\selectfont

\caption{{\small Divide-and-Conquer edit distance algorithm on $A[1..n]$ and $B[1..n]$.\label{algo:dac-mm}}}

\SetKw{Break}{break}
\SetKw{Kwand}{and}
\SetKw{Kwor}{or}
\SetKw{And}{and}
\SetKw{Or}{or}
\SetNlSkip{0.2em}

\SetKwInput{note}{Notes}


\DontPrintSemicolon
\fontsize{9pt}{9pt}
\selectfont

\begin{minipage}[t]{.47\columnwidth}
\note{
We assume both $A$ and $B$ has size $n=2^c$ for simple description.
Our algorithm also works for strings with different lengths with minor changes.}
\MyFunc{\upshape \mf{\dacmm{}}} {
$t \gets 1$ \;
\While{\true} {
    $\dist \gets \fname{Check{}}(t)$ \;
    \lIf {$\dist[t][t] \leq t$} {
        \Break    }
    $t \gets \min(2t, n)$ \;
}
\Return{$\dist[t][t]$}\;
}
\codeskip
\tcp{Find the SD in the DP matrix from $(0,0)$ to $(n,n)$ by restricting in the diagonal stripe with width $t$}
\MyFunc{\upshape \fname{Check}$(t)$}{
    $\dist \gets$\fname{\dacmmrec}$(0,0,n,t)$
}
\end{minipage}\hfill\begin{minipage}[t]{0.47\columnwidth}
\MyFunc{\upshape \fname{\dacmmrec{}}$(i,j,n,t)$} {
    \If {$n/2 < t$} {
        Computed $\dist $ by the \aalm{} algorithm \;
        \Return {$\dist$}
    }
    \tcp{Compute the SD matrices for $\dpv_1,\dpv_2,\dpv_3,\dpv_4$ (as shown in \cref{fig:dac_mm} (d)).}
    $\dist_1 \gets \mf{\dacmmrec{}}(i, j, n/2, t)$ \;
    Compute $\dist_2$ and $\dist_3$ by the \aalm{} algorithm \;
    $\dist_4 \gets \mf{\dacmmrec{}}(i+n/2, j+n/2, n-n/2, t)$ \;
    \tcp{use the same \fname{Combine} function as \aalm{}}
    $\dist_{1\cup 2} \gets \fname{Combine}(\dist_1, \dist_2)$ \;
    $\dist_{3\cup 4} \gets \fname{Combine}(\dist_3, \dist_4)$ \;
    $\dist \gets \fname(\dist_{1\cup 2}, \dist_{3\cup 4})$ \;
    \Return{$\dist$}\;
}
\end{minipage}
\end{algorithm}
However, the output value $k$ is unknown before we run the algorithm.
To overcome this issue, we use a strategy based on prefix doubling to ``binary search'' the value of $k$ without asymptotically increasing the work of the algorithm.
We start with $t=1$, 
and run the \fname{Check}$(t)$ in \cref{algo:dac-mm} (i.e., restricting the search in a width-$(2t+1)$ diagonal).
Assume that the \fname{Check} function returns $\sigma$ edits.
If $\sigma \leq t$, we know that $\sigma$ is the \SP{} from $(0,0)$ to $(n,n)$,
since allowing the path to go out of the diagonal area will result in an answer greater than $t$.
Otherwise, we know $\sigma > t$, and $\sigma$ is not necessarily the \shortestdis{} from the $(0,0)$ to $(n,n)$,
since not restricting the path in the $t$-diagonal area may allow for a shorter path.
If so, we double $t$ and retry.
Although we need $O(\log k)$ searches before finding the final answer $k$, we will show that the total search cost is asymptotically bounded by the last search.
In the last search, we have $t< 2k$.

We first analyze the cost for \fname{Check}$(t)$.
It contains two recursive calls, two calls to \aalm{}, and three calls to the \fname{Combine} function.
Therefore, the work for \fname{Check}$(t)$ is $W(n)=2W(n/2)+O(t^2\log t)$, with base cases $W(t)=t^2\log t$,
which solves to $W(n)=O(nt\log t)$.
For span, note that there are $\log (n/t)$ levels of recursion before reaching the base cases.
In each level, the \fname{Combine} function combines $t\times t$ \SP{} matrices with $O(\log^2 t)$ span.
In the leaf level, the base case uses \aalm{} with $O(\log^3 t)$ span. Therefore, the total span of a \fname{Check} is:
\begin{equation}
O(\log n/t \cdot \log^2 t + (\log n/t + \log^3 t)) = O(\log^2 t\cdot(\log n/t+\log t))=O(\log^2 t\log n)
\end{equation}
We will apply \fname{Check}$(\cdot)$ for $O(\log k)$ times, with $t=1,2, 4, \dots $ up to at most $2k$.
Therefore, the total work is dominated by the last \fname{Check}, which is $O(nk\log k)$. The span is $O(\log n\log^3 k)$. 
\begin{theorem}
    The \dacmm{} algorithm computes the edit distance between two sequences of length $n$ and $m\le n$ in $O(nk\log k)$ work and $O(\log n\log^3 k)$ span,
    where $k$ is the output size (fewest possible edits).
\end{theorem}
Compared to the BFS-based algorithms with $\tilde{O}(k)$ span, our \dacmm{} is also output-sensitive and achieves polylogarithmic span.
However, the work is $\tilde{O}(kn)$ instead of $\tilde{O}(n+k^2)$, which will lead to more running time in practice for a moderate size of $k$.

\section{Implementation Details}

We provide all implementations for the four algorithms as well as testing benchmarks at~\cite{edcode}.
In this section, we highlight some interesting and challenging parts of our implementations.

\subsection{Implementation Details of BFS-based Algorithms}\label{sec:impl-sa}

For the suffix array construction in \bfssa{},
we implemented a parallel version of the DC3 algorithm \cite{karkkainen2003simple}. 
We also compared our implementation with the SA implementation in ParlayLib~\cite{blelloch2020parlaylib},
which is a highly optimized version of the prefix doubling algorithm with $O(n\log n)$ work and $O(\log^2n)$ span.
On average, our implementation is about 2$\times$ faster than that in ParlayLib when applied to edit distance.
We present some results for their comparisons in \iffullversion{\cref{tab:sa_compare} in the appendix.}\ifconference{the full version~\cite{ding2023efficientfull}.}
For LCP array construction and preprocessing RMQ queries, we use the implementation in ParlayLib~\cite{blelloch2020parlaylib},
which requires $O(n\log n)$ work and $O(\log^2n)$ span.
With them, the query has $O(1)$ cost.

In our experiments on both synthetic and real-world data,
we observed that the LCP length is either very large when we find two long matched chunks,
or in most of the cases, very short when they are not corresponding to each other.
This is easy to understand---for genomes, text or code with certain edit history, it is unlikely that two random starting positions share a large common prefix.
Based on this, we add a simple optimization for all LCP implementations such that we first compare the leading eight characters,
and only when they all match, we use the regular LCP query.
This simple optimization greatly improved the performance of \bfssa{}, and also slightly improved the hash-based solutions.

\hide{
Interestingly, we observe that the standard LCP implementation on SA mentioned above is much slower than the hash-based LCP solutions.
\yihan{Is it still true?}
We take a careful study on this to understand the reason.
We note that for both synthetic and real-world data, the LCP returned is either very large when we find two matched chunks with no edits,
or in most of the cases, very short when they are not corresponding to each other.
This is easy to understand---for genomes, text or code with certain edit history, it is unlikely that two random starting positions share a large common prefix.
Hence, in most cases, the hash-based LCP finishes in a few steps in dual binary search.
Meanwhile, since most LCPs are small, most of the BFS visits (see \cref{fig:bfs}) are local and close to previously visited cells.
Hence, the check on Line \ref{line:lastcheck} is likely a cache hit and incurs a small cost.
In contrast, sorted by suffixes, $A[x..n]$ and $A[(x+1)..n]$ are likely to be separated in the SA, so even the LCP queries on nearby starting positions are likely to incur random memory accesses, which hampers the efficiency.
Thus, we add a simple optimization for the SA-based LCP query by first checking a block of size $b$, and if they all match, we run the regular RMQ-based LCP query.
This simple optimization significantly improves the performance of the SA-based LCP for all our tests by a factor of 5--10$\times$, and matches the performance of the hash-based solutions. \yihan{check the numbers}
We believe this simple yet efficient optimization can also be considered for other applications with such query distributions.
\yihan{may need to rewrite this paragraph after checking more details}
}




\subsection{Implementation Details of the \dacmm{} Algorithm}\label{sec:dac-imp}
Although our \dacmm{} algorithm given in \cref{algo:dac-mm} is not complicated, we note that implementing it is highly non-trivial
in two aspects.
First, in \cref{sec:dac-sp}, we assume both strings $A$ and $B$ have the same length $n$, which is a power of two.
However, handling two strings with different lengths makes the matrix partition more complicated in practice.
Another key challenge is that the combining step in the \aalm{} algorithm is recursive
and needs to allocate memory with varying sizes in the recursive execution.
While memory allocation is mostly ignored theoretically,
frequent allocation in practice can easily be the performance bottleneck in the parallel setting.
We discuss our engineering efforts as follows.

\vspace{-10px}
\subparagraph*{Irregularity.}
The general case, when $n$ and $m$ are not powers of two and not the same, is more complicated than the case in \cref{algo:dac-mm}.
In this case, all four subproblems $G_1$, $G_2$, $G_3$, and $G_4$ will have different sizes.
While theoretically, we can always round up, for better performance in practice,
we need to introduce additional parameters to restrict the search within the belt region as shown in \cref{fig:dac_mm_k}.
Therefore, we use two parameters $t_1$ and $t_2$, to denote the lengths of the diagonal area on each side.
We show an illustration in \cref{fig:dac_mm_k}(a) along with how to compute the subproblem sizes.
In extreme cases, $t_1$ or $t_2$ can degenerate to $0$, which results in three subproblems (\cref{fig:dac_mm_k}(b)).
In such cases, we will first merge $\dpv_2$ and $\dpv_4$, then merge $\dpv_1$ and $\dpv_{2\cup 4}$.

\hide{
Ideally, if $n=m$ and it is a power of two, the partition scheme would work perfectly.
However, in the general case that $n \neq m$, the side length of the diagonal area of
$\dpv_4$ could be different ($m_1+t-n_1$ and $n_1+t-m_1$ in \cref{fig:dac_mm_k} (c)).
Thus, in practice, we need two parameters, $t_1$ and $t_2$, to denote the length of the diagonal area on each side (\cref{fig:dac_mm_k} (d)).
In some extreme cases, one of the triangles can even degenerate, which results in three partitions in this case (\cref{fig:dac_mm_k} (e)).
In such case, we can first merge $\dpv_2$ and $\dpv_4$, then merge $\dpv_1$ and $\dpv_{2\cup 4}$.
}

\vspace{-10px}
\subparagraph*{The Combining Step.}
As mentioned in \cref{sec:dac-sp}, achieving an efficient combining step is highly non-trivial.
The straightforward solution to combine two matrices is to use the Floyd-Warshall algorithm~\cite{floyd1962algorithm},
but it incurs $O(n^3)$ work and will be a bottleneck.
The \aalm{} algorithm improves this step to $O(n^2)$ by taking advantage of the Monge property of the two matrices.
For page limit, we introduce the details of the combining algorithm \iffullversion{in \cref{app:aalm-combine-algo}.}\ifconference{in the full version~\cite{ding2023efficientfull}.}
However, the original AALM algorithm is based on divide-and-conquer and requires memory allocation for every recursive function call.
This is impractical as frequent parallel memory allocation is extremely inefficient.
To overcome this challenge, we redesign the recursive solution to an iterative solution,
such that we can preallocate the memory space before the combining step.
No dynamic memory allocation is involved during the computation.
We provide the details of this approach \iffullversion{in \cref{app:aalm-combine-impl}.}\ifconference{in the full version~\cite{ding2023efficientfull}.}

\section{Experiments}\label{sec:exp}
\useunder{\uline}{\ul}{}

\begin{figure}

\hspace{-.3in}\begin{minipage}[H]{0.44\textwidth}
\input{figures/DAC_MM_K_fig.tex}
\end{minipage}%
\hfill\begin{minipage}[H]{0.58\textwidth}

\begin{table}[H]

\vspace{-2.75em}
\centering\small
    \begin{tabular}{@{}c@{ }@{ }c@{ }@{ }@{ }@{ }c@{ }@{ }@{ }@{ }c@{ }@{ }@{ }@{ }c@{ }@{ }@{ }@{ }c@{ }}
        {\textbf{Data}} &
          {\textbf{Alias} } &
          {$\boldsymbol{|A|}$} &
          {$\boldsymbol{|B|}$} &
          {\textbf{$\boldsymbol{k}$}} &
          {$|\boldsymbol{\Sigma}|$} \\ \hline  
        \multirow{2}{*}{Wikipedia} &
          Wiki v1 &
          0.56M &
          0.56M &
          439 &
          256 \\ 
         \multirow{2}{*}{pages~\cite{WikiQuebec}}& Wiki v2 & 0.56M   & 0.56M   & 5578  & 256 \\ 
         & Wiki v3 & 0.56M   & 0.55M   & 15026 & 256 \\ \hline
        \multirow{2}{*}{Linux kernel} &
          Linux v1 &
          6.47M &
          6.47M &
          236 &
          256 \\ 
         \multirow{2}{*}{code~\cite{LinuxKernel}}& Linux v2    & 6.47M  & 6.47M  & 1447  & 256 \\ 
         & Linux v3    & 6.47M  & 6.46M  & 9559  & 256 \\ \hline
        \multirow{2}{*}{DNA} &
          DNA 1 &
          42.3M &
          42.3M &
          928 &
          4 \\ 
         \multirow{2}{*}{sequences~\cite{benson2012genbank}}& DNA 2    & 42.3M & 42.3M & 9162  & 4   \\ 
         & DNA 3    & 42.3M & 42.3M & 91419 & 4   \\ \hline
        \end{tabular}
        \vspace{-.5em}\caption{\small \textbf{Real-world datasets in our experiments, including input sizes $|A|$ and $|B|$, number of edits $k$, and alphabet sizes $|\Sigma|$.}\vspace{-1.5em}}
        \label{tab:datasets}
\end{table}

\end{minipage}
\vspace{-1.5em}
\end{figure} 

\subparagraph*{Setup.}
We implemented all algorithms in C++ using ParlayLib~\cite{blelloch2020parlaylib} for fork-join parallelism and some parallel primitives (e.g., reduce).
Our tests use a 96-core (192 hyperthreads) machine with four Intel Xeon Gold 6252 CPUs, and 1.5 TB of main memory.
We utilize \texttt{numactl -i all} in tests with
more than one thread to spread the memory pages across CPUs in a round-robin fashion.
We run each test three times and report the median.

\newcommand{\figureoffset}{\hspace{-2em}}
\begin{figure*}[t]
    \centering
    \small
    \vspace{-0.9em}
    \textbf{Synthetic Datasets:}\\
    \figureoffset\includegraphics[width=\columnwidth]{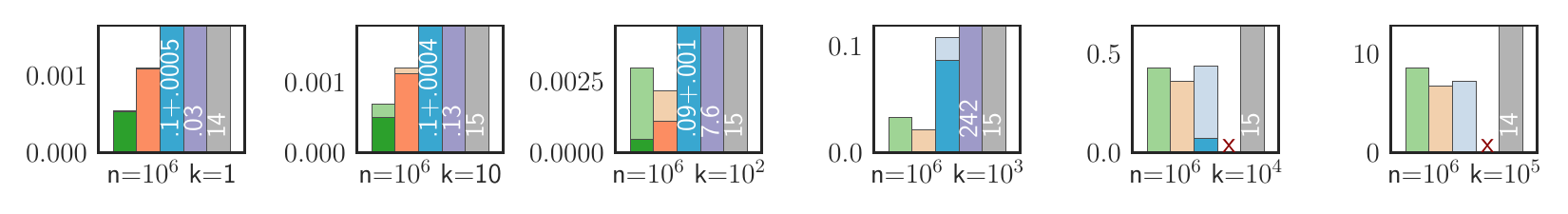}\\
    \vspace{-0.9em}
    \figureoffset\includegraphics[width=\columnwidth]{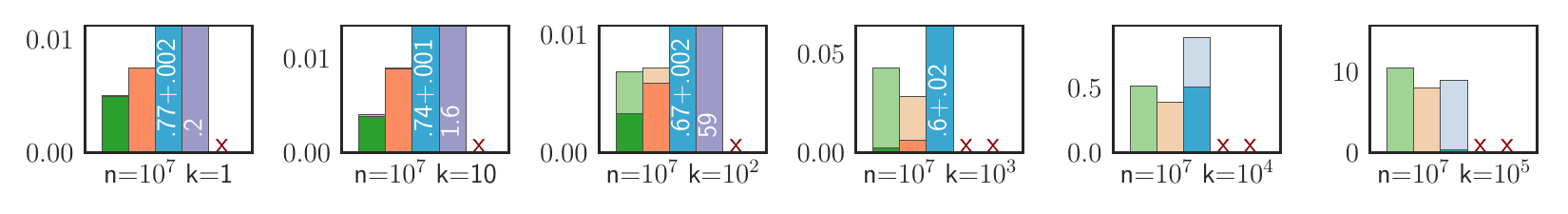}\\
    \vspace{-0.9em}
    \figureoffset\includegraphics[width=\columnwidth]{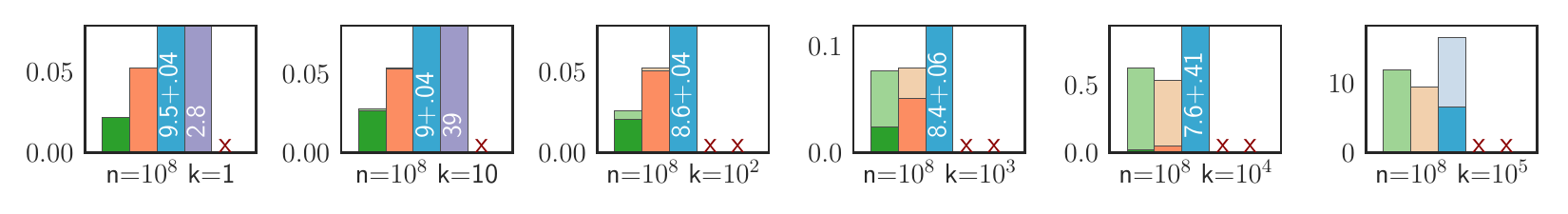}\\
    \vspace{-0.9em}
    \figureoffset\includegraphics[width=\columnwidth]{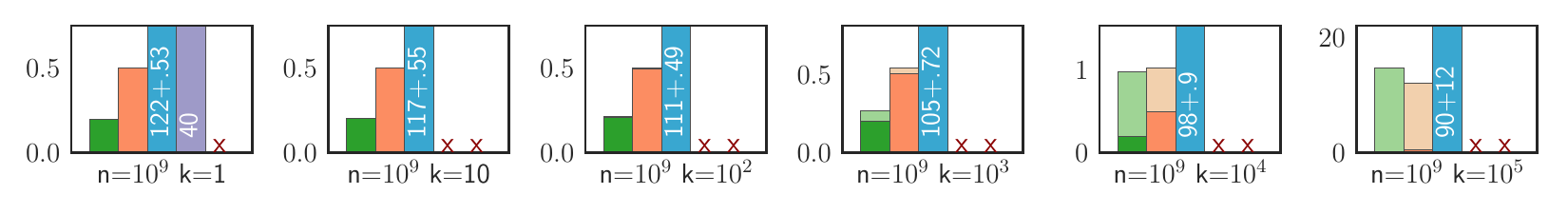}\\
    \vspace{-.5em}
    \textbf{Real-world Datasets:}\\
    \figureoffset\includegraphics[width=\columnwidth]{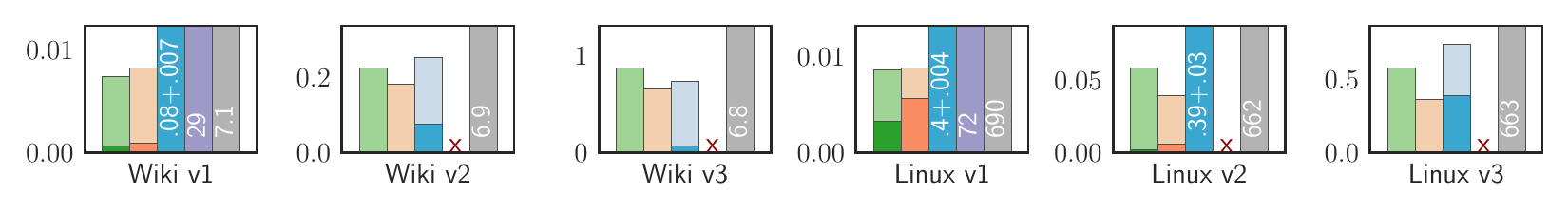}\\
    \vspace{-0.9em}
    \figureoffset\includegraphics[width=\columnwidth]{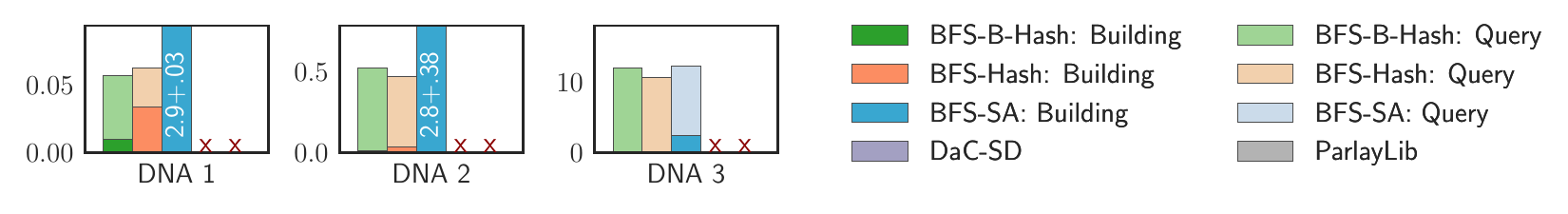}\\
    \vspace{-1.5em}
    \caption{\small \textbf{Running time (in seconds) of synthetic and real-world datasets for all algorithms.}
    Lower is better.
    We put an ``$\times$'' if the algorithm does not finish within 1000 seconds.
    For BFS-based algorithms, we separate the time into building time (constructing the data structure for LCP queries)
    and query time (running BFS).
    All bars out of the range of the y-axis are annotated with numbers.
    The number is the total running time for \dacmm{} and ParlayLib, and is in the format of $a+b$ for \bfssa{},
    where $a$ is the building time and $b$ is the query time.
    Full results are presented \iffullversion{in \cref{tab:full} in the appendix.}\ifconference{in the full version~\cite{ding2023efficientfull}.}
    \vspace{-2em}
    }
    \label{fig:time}
  \end{figure*} 

\begin{figure}

  \hspace{-.2in}
  \begin{minipage}[H]{0.62\textwidth}
    \vspace{-2em}
    \begin{figure}[H]
    \begin{subfigure}[b]{0.49\textwidth}
      \includegraphics[width=\textwidth]{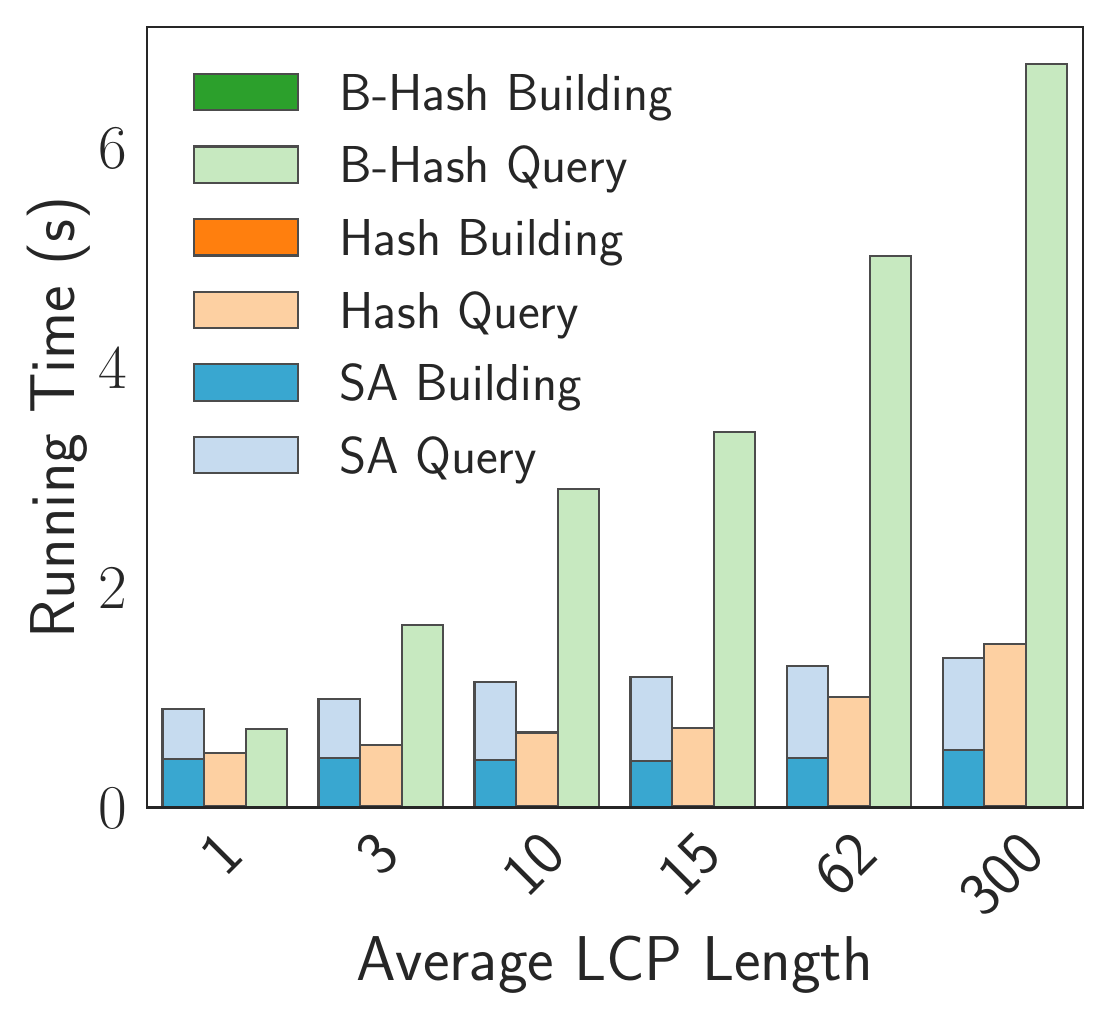}
      \vspace{-2em}
      \caption{\small\centering $n = 10^7$   $k=10000$}
    \end{subfigure}\hfill
    \begin{subfigure}[b]{0.5\textwidth}
      \includegraphics[width=\textwidth]{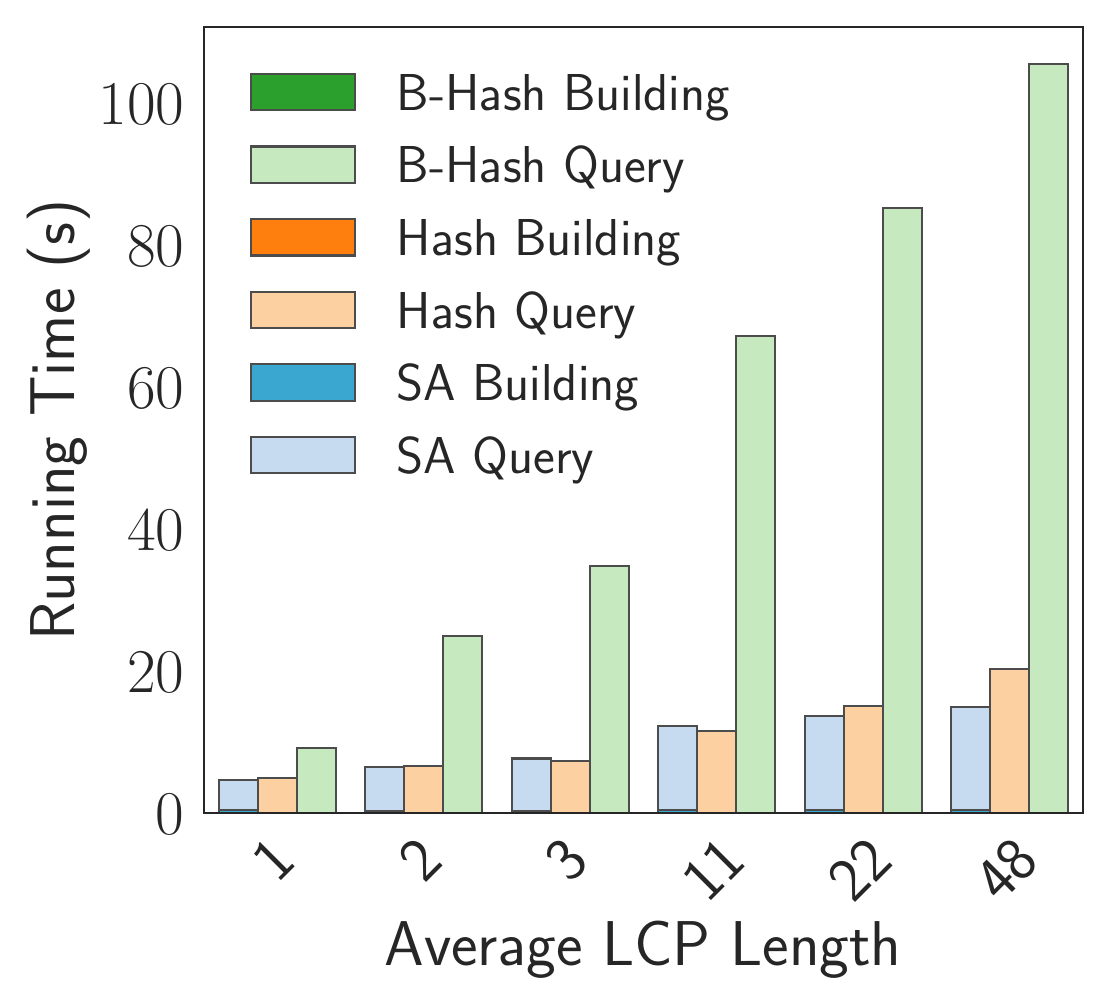}
      \vspace{-2em}
      \caption{\small\centering $n = 10^7$   $k=50000$}
    \end{subfigure}
      \vspace{-0.6em}
    \caption{\small\textbf{Performance of BFS-based algorithms vs. average LCP length.} Some building times are invisible because they are too small.\label{fig:lcp}}
  \end{figure}\hfill
  \vspace{-1em}
  \end{minipage}%
  \hfill
  \begin{minipage}[H]{0.36\textwidth}
    \vspace{-0.8em}
    \begin{figure}[H]
      \vspace{-1.5em}
      \centering
      \small
      \figureoffset
      \includegraphics[width=\columnwidth]{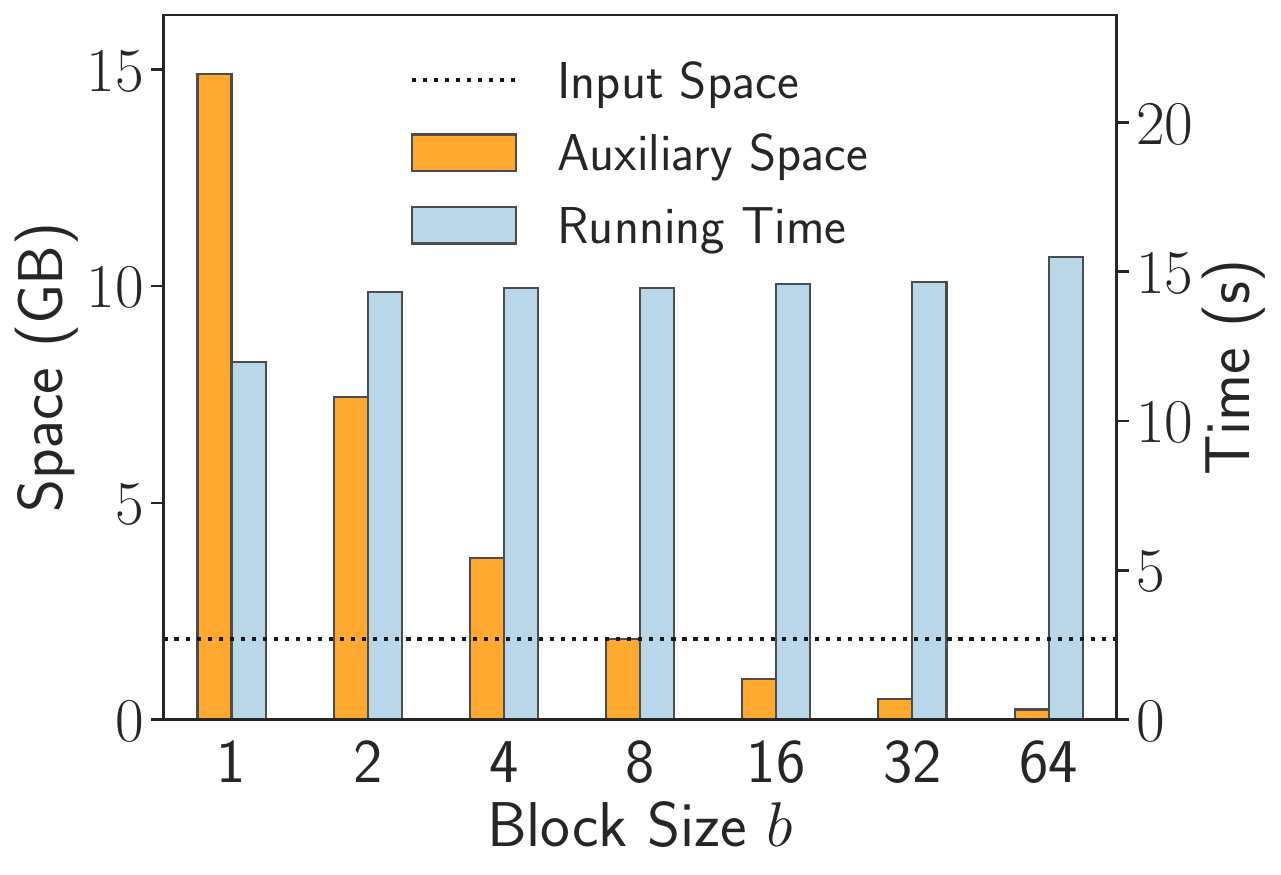}\\
      \vspace{-0.5em}
      $n=10^8$ $k=10^5$
      \vspace{-0.5em}
      \caption{\small\textbf{Time-space trade-off in \bfshash{}.}
      The space shown is the memory required for the \hashtab{}s. The dotted line is the input size. Note that by setting $b=1$, the algorithm is equivalent to \bfssimplehash{}.
      \label{fig:space}}
    \end{figure}
  \end{minipage}
  \vspace{-1.5em}
  \end{figure} 
\hide{
\begin{figure*}[t]
    \centering
    \small
    \figureoffset\includegraphics[width=0.6\columnwidth]{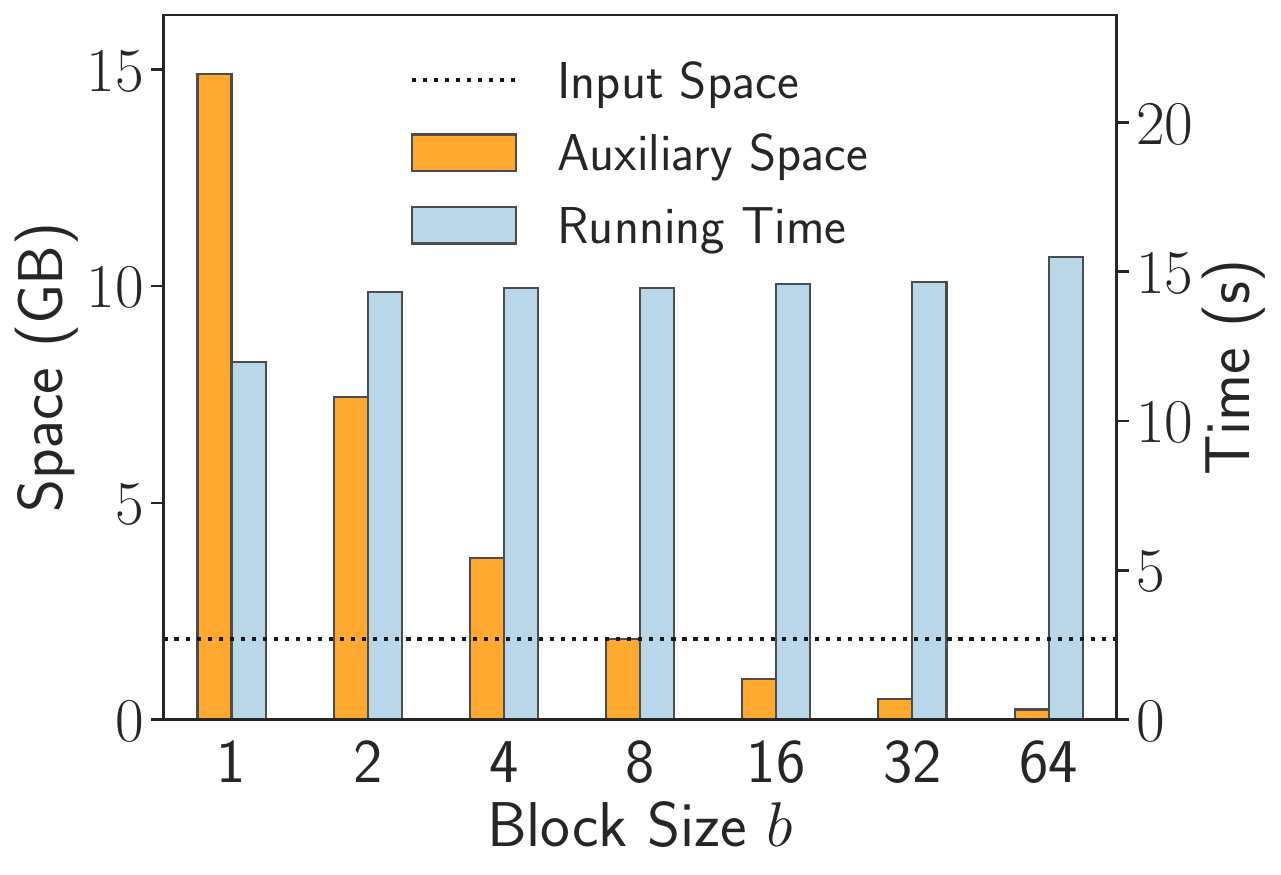}\\
    \vspace{-1.5em}
    \caption{\small \textbf{Compress ratio and running time for $n = 10^8, \#edits = 10^5$.  }
    \vspace{-2em}}
    \label{fig:compress}
  \end{figure*} 
}
\vspace{-10px}
\subparagraph*{Tested Algorithms and Datasets.}
We tested five algorithms in total:
four output-sensitive algorithms in this paper (\bfssa{}, \bfssimplehash{}, \bfshash{}, \dacmm{}),
and a baseline algorithm from ParlayLib~\cite{blelloch2020parlaylib}, which is a parallel output-insensitive implementation with $O(nm)$ work.
The ParlayLib implementation is intended to showcase the simplicity of parallel algorithms, and as a result, it may not be well-ptimized.
We are unaware of other parallel implementations that provide output-sensitive cost bounds.
We use $b=32$ for our \bfshash{}.
As we will show later, the running time is generally stable with $4\le b\le 64$.
We tested the algorithms on both synthetic and real-world datasets.
For synthetic datasets, we generate random strings with different string lengths $n=10^i$ for $6\le i\le 9$ and $k$ (number of edits) varying from 1 to $10^5$, and set the size of the alphabet as $256$.
We create strings $A$ and $B$ by generating $n$ random characters, and applying $k$ edits.
The $k$ edits are uniformly random for insertion, deletion and substitution.
For $k\ll n$, we have $m\approx n$.
All values of $k$ shown in the figures and tables are approximate values.
Our real-world datasets include Wikipedia~\cite{WikiQuebec}, Linux kernel~\cite{LinuxKernel}, and DNA sequences~\cite{peterson2009nih}.
We compare the edit distance between history pages on Wikipedia and history commits of a Linux kernel file on GitHub.
We also compare DNA sequences by adding valid modifications to them to simulate DNA damage or genome editing techniques, as is used in many existing papers~\cite{carroll2017focus,clement2019crispresso2, jeon2014ezeditor, xue2015ageseq}.
We present the statistics of the real-world datasets in \cref{tab:datasets}.


\vspace{-10px}
\subparagraph*{Overall Performance on Synthetic Data.}
We present our results on synthetic data in the upper part of \cref{fig:time}.
We also present the complete results \iffullversion{in \cref{tab:full}.}\ifconference{in the full version~\cite{ding2023efficientfull}.}
For BFS-based algorithms, we also separate the time for \defn{building} the data structures for LCP queries, and the \defn{query} time (the BFS process).
ParlayLib cannot process instances with $n>10^6$ due to its $O(nm)$ work bound.

We first \emph{compare our solutions with ParlayLib}~\cite{blelloch2020parlaylib}.
Since ParlayLib is not output-sensitive, its running time remains the same regardless of the value of $k$.
Among the tests that ParlayLib can process ($n=10^6$), our output-sensitive algorithms are
much faster than ParlayLib, especially when $k$ is small (up to $10^5 \times$).
For $n=10^6$, all our BFS-based algorithms are at least $1.7\times$ faster than ParlayLib even when $k\approx n/10$.

We then \emph{compare our DaC- and BFS-based solutions}.  \dacmm{} has the benefit of polylogarithmic span, compared to $\tilde{O}(k)$ span for the BFS-based algorithm.
Although this seems to suggest that \dacmm{} should have better performance when $k$ is large,
the result shows the opposite. The reason is that \dacmm{} has $\tilde{O}(nk)$ work,
compared to $\tilde{O}(n+k^2)$ cost of the BFS-based algorithms.
When $k$ becomes larger, the overhead in work is also more significant.
On the other hand, when $k$ is small, the $O(nk)$ work becomes linear,
which hides the inefficiency in work. Therefore, the gap between \dacmm{} and other algorithms is smaller when $k$ is small,
but \dacmm{} is still slower than BFS-based algorithms in all test cases, especially when $k$ is large.
This experiment reaffirms the \emph{importance of work efficiency} on practical performance for parallel algorithms.

Finally, we \emph{compare all our BFS-based solutions}.
Our hash-based solutions have significant advantages over the other implementations when $k$ is small, since the pre-processing time for hash-based solutions is much shorter.
When $k$ is large, pre-processing time becomes negligible, and \bfssimplehash{} seems to be the ideal choice since its query is also efficient.
In particular, for $n\approx m\approx10^9$, hash-based algorithms use about $1$ second for pre-processing while \bfssa{} uses about 100 seconds.
Although \bfssa{} also has $O(n)$ construction time, the constant is much larger and its memory access pattern is much worse than the two hash-based solutions.
We note that in some cases, the query time of \bfssa{} can still be faster than \bfssimplehash{} and \bfshash{}, especially when $k$ is large,
which is consistent with the theory ($O(1)$ vs. $O(\log n)$ or $O(b\log n)$ per LCP query).

In theory, \bfshash{} reduces space usage in \bfssimplehash{} by increasing the query time.
Interestingly, when $k$ is small, \bfshash{} can also be faster than \bfssimplehash{} by up to 2.5$\times$.
This is because \bfshash{} incurs fewer writes (and thus smaller memory footprints) in preprocessing that leads to faster building time.
When $k$ is small, the running time is mostly dominated by the building time, and thus \bfshash{} can perform better.
When $k$ is relatively large and $k^2$ is comparable to $n$,
\bfssimplehash{} becomes faster than \bfshash{} due to better LCP efficiency.
In fact, when $k$ is large, the running time is mainly dominated by the query (BFS), and
all three algorithms behave similarly.
It is worth noting that in these experiments with $|\Sigma|=256$ and random edits,
in most of the cases, the queried LCP is small.
Therefore, the $O(\log n)$ or $O(b\log n)$ query time for \bfssimplehash{} and \bfshash{} are not tight, and they have much better memory access patterns than \bfssa{}
in LCP queries.
As a result, they can have matching or even better performance than \bfssa{}.
Later we will show that under certain input distributions where the average LCP length is large,
\bfssa{} can have some advantage over both \bfssimplehash{} and \bfshash{}.

\vspace{-12px}
\subparagraph*{Real-World Datasets.} We now analyze how our algorithms
perform on real-world string and edit patterns.
The results are shown in the lower part of \cref{fig:time}.
The results are mostly consistent with our synthetic datasets, where \bfshash{} is more advantageous when $k$ is small,
and \bfssimplehash{} performs the best when $k$ is large.
When $k$ is large, \bfssa{} can also have comparable performance to the hash-based solutions.
\hide{
Especially when $k$ is large (more LCP queries are performed), the advantage of \bfssa{} becomes more significant.
For the same reason, \bfshash{} may perform much worse, since its overhead in query can be more pronounced.
Inspired by this, we further construct more skewed data to study how the average LCP length affect running time.
}

\hide{
Overall, \bfshash{} is the best across all different tests.
The benefit of \bfshash{} over \bfssimplehash{} is the more efficient preprocessing,
which has $O(n)$ work compared to $O(n\log n)$ for \bfssimplehash{}.
Indeed from \cref{fig:time}, we can see that when $k$ is small where the building cost dominates the time,
\bfshash{} has a more significant benefit than \bfssimplehash{}.
When $k$ becomes larger, the preprocessing time becomes negligible.
In this case, \bfssimplehash{} can be slightly faster since its query computation is simpler than \bfshash{}.
}

\hide{
The benefit of \bfshash{} over \bfssa{} also lies in better preprocessing time.
Through the correspondence with the authors, we believe the SA implementation we used is already the state-of-the-art, with $O(n\log n)$ work in the worst case but is fast for most real-world instances.
However, the hidden constant in the cost bounds is still large, since the algorithm involves multiple rounds of global shuffle with inefficient memory access patterns.
Hence, the building time for \bfssa{} is 15--62$\times$ slower than \bfssimplehash{} and 60--460$\times$ slower than \bfshash{}.
This leads to a significant performance gap, especially when $k$ is small.
}

\vspace{-12px}
\subparagraph*{LCP Length vs. Performance.}
It seems that for both synthetic and real-world data shown above, our hash-based solutions are always better than \bfssa{}.
It is worth asking, whether \bfssa{} can give the best performance in certain cases, given that it has the best theoretical bounds (see \cref{tab:algo}).
By investigating the bounds carefully, \bfssa{} has better LCP query cost as $O(1)$, while the costs for \bfssimplehash{} and \bfshash{} are $O(\log L)$ and $O(b\log L)$, respectively, where $L$ is the LCP length.
This indicates that \bfssa{} should be advantageous when $k$ and $L$ are both large.
To verify this, we artificially created input instances with medium to large values of~$k$ and controlled average LCP query lengths, and showed the results in \cref{fig:lcp} on two specific settings.

\hide{
One of the reason is that although \bfssa{} may have better query time,
this advantage is only useful when $k$ is large (i.e., building time is negligible).
However, a large $k$ may also imply a smaller LCP length, which is advantageous for the hash-based solutions,
since their worst case query bound ($O(\log n)$ or $O(b\log n)$ is not tight.
To better understand how the performance of the BFS-based algorithms is affected by the LCP length,
we constructed datasets with the same edit distance $k$ but varying average length of LCP queries performed in the algorithm.
Not that for the same input sequence, the three BFS-based algorithms perform exactly the same LCP queries.
We fix $|\Sigma|=2$ and use skewed distributions of the two characters so that it easily incurs long LCPs.
We present two settings in Figure \ref{fig:lcp}.
}

The experimental result is consistent with the theoretical analysis.
The running time for \bfssimplehash{} increases slowly with $L$, while the performance of \bfshash{} grows \emph{much faster}, since it is affected by a factor of $O(b)$ more than \bfssimplehash{}.
The query time for \bfssa{} almost stays the same, but also increases slightly with increasing $L$.
This is because in general, with increasing $L$,
the running time for all three algorithms may increase slightly due to worse cache locality in BFS due to more long matches.
In Figure \ref{fig:lcp}(a), the building time for both \bfssimplehash{} and \bfshash{} are negligible,
while \bfssa{} still incurs significant building time.
Even in this case, with an LCP length of 300, the query time of the hash-based solutions still becomes larger than the \emph{total} running time of \bfssa{}.
In Figure \ref{fig:lcp}(b) with a larger $k$, the building time for all three algorithms is negligible. In this case, \bfssa{} always has comparable performance with \bfssimplehash{}, and may perform better when $L>20$.
However, such extreme cases (both $k$ and $L$ are large) should be very rare in real-world datasets -
when $k$ is large enough so that the query time is large enough to hide SA's building time,
$L$ is more likely to be small,
which in turn is beneficial for the query bounds in hash-based solutions.
Indeed such cases did not appear in our 33 tests on both synthetic and real data.

\vspace{-15px}
\subparagraph*{Parallelism.} We test the self-relative speedup of all algorithms. We present speedup numbers on two representative tests with different values of $n$ and $k$ in \cref{fig:speedup}.
For BFS-based algorithms, we separate the speedup for building and query.
All our algorithms are highly parallelized.
Even though \bfssa{} and \dacmm{} have a longer running time, they still have a 48--68$\times$ speedup, indicating good scalability.
Our \bfssimplehash{} algorithm has about 40--50$\times$ speedup in building, and \bfshash{} has a lower but decent speedup of about 20--40$\times$.
When $k$ is small, the frontier sizes (and the total work) of BFS are small, and the running time is also negligible.
In this case, we cannot observe meaningful speedup.
For larger $k=10^5$, three BFS-based algorithms achieve 27--48$\times$ speedup both in query and entire edit distance algorithm.

\begin{table}[t]
    \centering
    \small
    \setlength{\tabcolsep}{2.6pt}
    \vspace{-.5em}
\begin{tabular}{ll|ccc|ccc|ccc|c}
    \multicolumn{1}{c}{\multirow{2}[1]{*}{\boldmath{}\textbf{$\boldsymbol{n}$}\unboldmath{}}} & \multicolumn{1}{c|}{\multirow{2}[1]{*}{\boldmath{}\textbf{$\boldsymbol{k}$}\unboldmath{}}} & \multicolumn{3}{c|}{\textbf{\bfshash{}}} & \multicolumn{3}{c|}{\textbf{\bfssimplehash{}}} & \multicolumn{3}{c|}{\textbf{\bfssa{}}} & \textbf{\dacmm{}} \\
          &       & \textbf{Build} & \textbf{Query} & \textbf{Total} & \textbf{Build} & \textbf{Query} & \textbf{Total} & \textbf{Build} & \textbf{Query} & \textbf{Total} & \textbf{Total} \\
    \midrule
    \boldmath{}\textbf{$10^8$}\unboldmath{} & \boldmath{}\textbf{$10$}\unboldmath{} & 20.4  & -     & 19.9  & 46.6 & -     & 46.5  & 49.6  & -     & 49.4 & 68.2\\
    \boldmath{}\textbf{$10^9$}\unboldmath{} & \boldmath{}\textbf{$10^5$}\unboldmath{} & 24.2 & 36.4  & 36.3  & 42.7  & 46.8  & 46.6 & 51.2  & 27.1  & 48.3  & t.o. \\
    \end{tabular}%
    \vspace{-.5em}
    \caption{\small \textbf{Self-relative speedup of each implementation in each step.}
    ``Build'' $=$ constructing the data structure for LCP queries.
    ``Query'' $=$ the BFS process.
    ``t.o.'' $=$ timeout.
    We omit query speedup when $k=10$ because there is little parallelism to be explored for BFS with small $k$,
    and the BFS time is also small and hardly affects the overall speedup. 192 hyperthreads are used for parallel executions.
    \vspace{-1.5em}
    }
    \label{fig:speedup}
\end{table}%

\vspace{-12px}
\subparagraph*{Space Usage.} We study the time-space tradeoff of our \bfshash{} with different block sizes $b$.
We present the \emph{auxiliary space} used by the \hashtab{} in \bfshash{} along with running time in \cref{fig:space} using one test case with $n=10^8$ and $k=10^5$ in our synthetic dataset.
The dotted line shows the input size.
Note that when $b=1$, it is exactly \bfssimplehash{}.
Since the inputs are 8-bit characters and the hash values are 64-bit integers, \bfssimplehash{} incurs $8\times$ space overhead than the input size.
Using blocking, we can avoid such overhead and keep the auxiliary space even lower than the input.
The auxiliary space decreases linearly with the block size $b$.
Interestingly, although blocking itself incurs time overhead, the impact in time is small: the time grows by $1.19 \times$ from $b=1$ to $2$,
and grows by $1.08 \times$ from $b=2$ to $64$.
This is mostly due to two reasons: 1) as mentioned, with 8-bit character input type and random edits, the average LCP length is likely short and within the first block,
and therefore the query costs in both approaches are close to $O(L)$ for LCP length $L$, and
2) the extra factor of $b$ in queries (Line \ref{line:lastcheck}) is
mostly cache hits (consecutive locations in an array).
This illustrates the benefit of using blocking in such datasets, since blocking saves much space while only increasing the time by a small fraction.

\hide{Since these datasets are from real edit history of the same document (Wikipedia page or GitHub code commit),
their edit distances are generally small.
\bfshash{} is significantly faster than other algorithms in these cases when the edit distance is small.
However, with the decreasing of the edit distance, the increase of the overall time of \bfssa is slower due to the better query performace.
We also test some test cases where the two versions are apart by a long time or significant changes happen to a sequence, and the edits are large (e.g., Wiki v3 and DNA Pair 3).
For the computing of edit distance of the DNA pair 3, SA performs the best among all the BFS-baed algotihms.
}


\section{Conclusion and Discussions}

We proposed output-sensitive parallel algorithms for the edit-distance problem, as well as careful engineering of them.
We revisited the BFS-based Landau-Vishkin algorithm.
In addition to using SA as is used in Landau-Vishkin (our \bfssa{} implementation),
we also designed two hash-based data structures to replace the SA for more practical LCP queries (\bfssimplehash{} and \bfshash{}).
We also presented the first output-sensitive parallel algorithm based on divide-and-conquer with $\tilde{O}(nk)$ work and polylogarithmic span.
We have also shown the best of our engineering effort on this algorithm, although its performance seems less competitive than other candidates due to work inefficiency.

We implemented all these algorithms and tested them on synthetic and real-world datasets. 
In summary, our BFS-based solutions show the best overall performance on datasets with real-world edits or random edits,
due to faster preprocessing time and better I/O-friendliness.
\bfssimplehash{} performs the best in time when $k$ is large.
\bfshash{} has better performance when $k$ is small.
The blocking scheme also greatly improves space efficiency without introducing much overhead in time.
In very extreme cases where both $k$ and the LCP lengths are large, \bfssa{} can have some advantages over the hash-based solutions,
while \bfshash{} can be much slower than \bfssimplehash{}.
However, such input patterns seem rare in the real world.

All our BFS-based solutions perform better than the output-insensitive solution in ParlayLib,
and the DaC-based solution with $\tilde{O}(nk)$ work and polylogarithmic span, even for large $k>\sqrt{n}$.
The results also imply the importance of work efficiency in parallel algorithm designs, consistent with the common belief in the literature~\cite{shen2022many,gu2023parallel}.
Because the number of cores in modern multi-core machines is small (usually hundreds to thousands) compared to the problem size,
an algorithm is less practical if it blows up the work significantly, as parallelism cannot compensate for the performance loss
due to larger work.



\clearpage

\begin{thebibliography}{10}

\bibitem{apostolico1990efficient}
Alberto Apostolico, Mikhail~J Atallah, Lawrence~L Larmore, and Scott McFaddin.
\newblock Efficient parallel algorithms for string editing and related
  problems.
\newblock {\em {SIAM} J. on Computing}, 19(5):968--988, 1990.

\bibitem{arora2001thread}
Nimar~S Arora, Robert~D Blumofe, and C~Greg Plaxton.
\newblock Thread scheduling for multiprogrammed multiprocessors.
\newblock {\em Theory of Computing Systems (TOCS)}, 34(2):115--144, 2001.

\bibitem{babu1997parallel}
K~Nandan Babu and Sanjeev Saxena.
\newblock Parallel algorithms for the longest common subsequence problem.
\newblock In {\em {IEEE} International Conference on High Performance Computing
  (HiPC)}, pages 120--125. IEEE, 1997.

\bibitem{bender2000lca}
Michael~A. Bender and Martin Farach-Colton.
\newblock The lca problem revisited.
\newblock In {\em Latin American Symposium on Theoretical Informatics (LATIN)},
  pages 88--94. Springer, 2000.

\bibitem{benson2012genbank}
Dennis~A Benson, Mark Cavanaugh, Karen Clark, Ilene Karsch-Mizrachi, David~J
  Lipman, James Ostell, and Eric~W Sayers.
\newblock Genbank.
\newblock {\em Nucleic acids research}, 41(D1):D36--D42, 2012.

\bibitem{Blelloch89}
Guy~E. Blelloch.
\newblock Scans as primitive parallel operations.
\newblock {\em IEEE Trans. on Comput.}, 38(11), 1989.

\bibitem{blelloch2020parlaylib}
Guy~E. Blelloch, Daniel Anderson, and Laxman Dhulipala.
\newblock Parlaylib --- a toolkit for parallel algorithms on shared-memory
  multicore machines.
\newblock In {\em {ACM} Symposium on Parallelism in Algorithms and
  Architectures (SPAA)}, pages 507--509, 2020.

\bibitem{blelloch2020optimal}
Guy~E. Blelloch, Jeremy~T. Fineman, Yan Gu, and Yihan Sun.
\newblock Optimal parallel algorithms in the binary-forking model.
\newblock In {\em {ACM} Symposium on Parallelism in Algorithms and
  Architectures (SPAA)}, pages 89--102, 2020.

\bibitem{BL98}
Robert~D. Blumofe and Charles~E. Leiserson.
\newblock Space-efficient scheduling of multithreaded computations.
\newblock {\em {SIAM} J. on Computing}, 27(1), 1998.

\bibitem{boucher2022bad}
Nicholas Boucher, Ilia Shumailov, Ross Anderson, and Nicolas Papernot.
\newblock Bad characters: Imperceptible nlp attacks.
\newblock In {\em IEEE Symposium on Security and Privacy (SP)}, pages
  1987--2004. IEEE, 2022.

\bibitem{carroll2017focus}
Dana Carroll.
\newblock Focus: genome editing: genome editing: past, present, and future.
\newblock {\em The Yale journal of biology and medicine}, 90(4):653, 2017.

\bibitem{cheon2015homomorphic}
Jung~Hee Cheon, Miran Kim, and Kristin Lauter.
\newblock Homomorphic computation of edit distance.
\newblock In {\em International Conference on Financial Cryptography and Data
  Security}, pages 194--212. Springer, 2015.

\bibitem{clement2019crispresso2}
Kendell Clement, Holly Rees, Matthew~C Canver, Jason~M Gehrke, Rick Farouni,
  Jonathan~Y Hsu, Mitchel~A Cole, David~R Liu, J~Keith Joung, Daniel~E Bauer,
  et~al.
\newblock Crispresso2 provides accurate and rapid genome editing sequence
  analysis.
\newblock {\em Nature biotechnology}, 37(3):224--226, 2019.

\bibitem{CLRS}
Thomas~H. Cormen, Charles~E. Leiserson, Ronald~L. Rivest, and Clifford Stein.
\newblock {\em Introduction to Algorithms (3rd edition)}.
\newblock MIT Press, 2009.

\bibitem{dasgupta2008algorithms}
Sanjoy Dasgupta, Christos~H Papadimitriou, and Umesh~Virkumar Vazirani.
\newblock {\em Algorithms}.
\newblock McGraw-Hill Higher Education New York, 2008.

\bibitem{edcode}
Xiangyun Ding, Xiaojun Dong, Yan Gu, Yihan Sun, and Youzhe Liu.
\newblock Parallel implementations for output-sensitive edit distance.
\newblock \url{https://github.com/ucrparlay/Edit-Distance}, 2023.

\bibitem{floyd1962algorithm}
Robert~W Floyd.
\newblock Algorithm 97: shortest path.
\newblock {\em Commun. {ACM}}, 5(6):345, 1962.

\bibitem{galil1986improved}
Zvi Galil and Raffaele Giancarlo.
\newblock Improved string matching with $k$ mismatches.
\newblock {\em ACM SIGACT News}, 17(4):52--54, 1986.

\bibitem{galil1987parallel}
Zvi Galil and Raffaele Giancarlo.
\newblock Parallel string matching with $k$ mismatches.
\newblock {\em Theoretical Computer Science (TCS)}, 51(3):341--348, 1987.

\bibitem{galil1988data}
Zvi Galil and Raffaele Giancarlo.
\newblock Data structures and algorithms for approximate string matching.
\newblock {\em Journal of Complexity}, 4(1):33--72, 1988.

\bibitem{galil1990improved}
Zvi Galil and Kunsoo Park.
\newblock An improved algorithm for approximate string matching.
\newblock {\em SIAM Journal on Computing}, 19(6):989--999, 1990.

\bibitem{goodrich2015algorithm}
Michael~T Goodrich and Roberto Tamassia.
\newblock {\em Algorithm design and applications}.
\newblock Wiley Hoboken, 2015.

\bibitem{gu2023parallel}
Yan Gu, Ziyang Men, Zheqi Shen, Yihan Sun, and Zijin Wan.
\newblock Parallel longest increasing subsequence and van emde boas trees.
\newblock In {\em {ACM} Symposium on Parallelism in Algorithms and
  Architectures (SPAA)}, 2023.

\bibitem{gu2022analysis}
Yan Gu, Zachary Napier, and Yihan Sun.
\newblock Analysis of work-stealing and parallel cache complexity.
\newblock In {\em {SIAM} Symposium on Algorithmic Principles of Computer
  Systems (APOCS)}, pages 46--60. SIAM, 2022.

\bibitem{harel1984fast}
Dov Harel and Robert~Endre Tarjan.
\newblock Fast algorithms for finding nearest common ancestors.
\newblock {\em siam Journal on Computing}, 13(2):338--355, 1984.

\bibitem{hirschberg1975linear}
Daniel~S. Hirschberg.
\newblock A linear space algorithm for computing maximal common subsequences.
\newblock {\em Commun. {ACM}}, 18(6):341--343, 1975.

\bibitem{hladek2020survey}
Daniel Hl{\'a}dek, J{\'a}n Sta{\v{s}}, and Mat{\'u}{\v{s}} Pleva.
\newblock Survey of automatic spelling correction.
\newblock {\em Electronics}, 9(10):1670, 2020.

\bibitem{hossain2019auto}
Md~Mosabbir Hossain, Md~Farhan Labib, Ahmed~Sady Rifat, Amit~Kumar Das, and
  Monira Mukta.
\newblock Auto-correction of english to bengali transliteration system using
  levenshtein distance.
\newblock In {\em International Conference on Smart Computing \& Communications
  (ICSCC)}, pages 1--5. IEEE, 2019.

\bibitem{jeon2014ezeditor}
Yoon-Seong Jeon, Kihyun Lee, Sang-Cheol Park, Bong-Soo Kim, Yong-Joon Cho,
  Sung-Min Ha, and Jongsik Chun.
\newblock Ezeditor: a versatile sequence alignment editor for both rrna-and
  protein-coding genes.
\newblock {\em International journal of systematic and evolutionary
  microbiology}, 64(Pt\_2):689--691, 2014.

\bibitem{jiang2002general}
Tao Jiang, Guohui Lin, Bin Ma, and Kaizhong Zhang.
\newblock A general edit distance between {RNA} structures.
\newblock {\em Journal of Computational Biology}, 9(2):371--388, 2002.

\bibitem{karkkainen2003simple}
Juha K{\"a}rkk{\"a}inen and Peter Sanders.
\newblock Simple linear work suffix array construction.
\newblock In {\em Intl. Colloq. on Automata, Languages and Programming
  {(ICALP)}}, pages 943--955. Springer, 2003.

\bibitem{karp1987efficient}
Richard~M Karp and Michael~O Rabin.
\newblock Efficient randomized pattern-matching algorithms.
\newblock {\em IBM journal of research and development}, 31(2):249--260, 1987.

\bibitem{krusche2010new}
Peter Krusche and Alexander Tiskin.
\newblock New algorithms for efficient parallel string comparison.
\newblock In {\em {ACM} Symposium on Parallelism in Algorithms and
  Architectures (SPAA)}, pages 209--216, 2010.

\bibitem{landau1986efficient}
Gad~M Landau and Uzi Vishkin.
\newblock Efficient string matching with $k$ mismatches.
\newblock {\em Theoretical Computer Science (TCS)}, 43:239--249, 1986.

\bibitem{landau1988fast}
Gad~M Landau and Uzi Vishkin.
\newblock Fast string matching with $k$ differences.
\newblock {\em J. Computer and System Sciences}, 37(1):63--78, 1988.

\bibitem{landau1989fast}
Gad~M Landau and Uzi Vishkin.
\newblock Fast parallel and serial approximate string matching.
\newblock {\em J. Algorithms}, 10(2):157--169, 1989.

\bibitem{levenshtein1966binary}
VI~Levenshtein.
\newblock Binary codes capable of correcting deletions, insertions and
  reversals.
\newblock In {\em Soviet Physics Doklady}, volume~10, page 707, 1966.

\bibitem{li2010survey}
Heng Li and Nils Homer.
\newblock A survey of sequence alignment algorithms for next-generation
  sequencing.
\newblock {\em Briefings in bioinformatics}, 11(5):473--483, 2010.

\bibitem{LinuxKernel}
{Linux Kernel File dcn\_1\_0\_sh\_mask.h Commit History on GitHub}.
\newblock
  \url{https://github.com/torvalds/linux/blob/master/drivers/gpu/drm/amd/include/asic_reg/dcn/dcn_1_0_sh_mask.h}.

\bibitem{lu1994parallel}
Mi~Lu and Hua Lin.
\newblock Parallel algorithms for the longest common subsequence problem.
\newblock {\em {IEEE} Transactions on Parallel and Distributed Systems},
  5(8):835--848, 1994.

\bibitem{manber1993suffix}
Udi Manber and Gene Myers.
\newblock Suffix arrays: a new method for on-line string searches.
\newblock {\em {SIAM} J. on Computing}, 22(5):935--948, 1993.

\bibitem{marccais2019locality}
Guillaume Mar{\c{c}}ais, Dan DeBlasio, Prashant Pandey, and Carl Kingsford.
\newblock Locality-sensitive hashing for the edit distance.
\newblock {\em Bioinformatics}, 35(14):i127--i135, 2019.

\bibitem{mccauley:LIPIcs.ICDT.2021.21}
Samuel McCauley.
\newblock {Approximate Similarity Search Under Edit Distance Using
  Locality-Sensitive Hashing}.
\newblock In {\em 24th International Conference on Database Theory (ICDT
  2021)}, volume 186, pages 21:1--21:22. Schloss Dagstuhl -- Leibniz-Zentrum
  f{\"u}r Informatik, 2021.

\bibitem{WikiQuebec}
{Municipal history of Quebec on Wikipedia}.
\newblock \url{https://en.wikipedia.org/wiki/Municipal_history_of_Quebec}.

\bibitem{myers1986nd}
Eugene~W Myers.
\newblock An o (nd) difference algorithm and its variations.
\newblock {\em Algorithmica}, 1(1-4):251--266, 1986.

\bibitem{myers1986incremental}
Eugene~Wimberly Myers.
\newblock {\em Incremental alignment algorithms and their applications}.
\newblock University of Arizona, Department of Computer Science, 1986.

\bibitem{myoupo1999time}
Jean-Fr{\'e}d{\'e}ric Myoupo and David Seme.
\newblock Time-efficient parallel algorithms for the longest common subsequence
  and related problems.
\newblock {\em J. Parallel Distrib. Comput.}, 57(2):212--223, 1999.

\bibitem{navarro2001guided}
Gonzalo Navarro.
\newblock A guided tour to approximate string matching.
\newblock {\em ACM computing surveys (CSUR)}, 33(1):31--88, 2001.

\bibitem{paterson1994longest}
Mike Paterson and Vlado Dan{\v{c}}{\'\i}k.
\newblock Longest common subsequences.
\newblock In {\em International Symposium on Mathematical Foundations of
  Computer Science}, pages 127--142. Springer, 1994.

\bibitem{peterson2009nih}
Jane Peterson, Susan Garges, Maria Giovanni, Pamela McInnes, Lu~Wang, Jeffery~A
  Schloss, Vivien Bonazzi, Jean~E McEwen, Kris~A Wetterstrand, Carolyn Deal,
  et~al.
\newblock The nih human microbiome project.
\newblock {\em Genome research}, 19(12):2317--2323, 2009.

\bibitem{shen2022many}
Zheqi Shen, Zijin Wan, Yan Gu, and Yihan Sun.
\newblock Many sequential iterative algorithms can be parallel and (nearly)
  work-efficient.
\newblock In {\em {ACM} Symposium on Parallelism in Algorithms and
  Architectures (SPAA)}, 2022.

\bibitem{shun2014fast}
Julian Shun.
\newblock Fast parallel computation of longest common prefixes.
\newblock In {\em International Conference for High Performance Computing,
  Networking, Storage, and Analysis (SC)}, pages 387--398. IEEE, 2014.

\bibitem{spinellis2012git}
Diomidis Spinellis.
\newblock Git.
\newblock {\em IEEE software}, 29(3):100--101, 2012.

\bibitem{tchendji2020efficient}
Vianney~Kengne Tchendji, Armel~Nkonjoh Ngomade, Jerry~Lacmou Zeutouo, and
  Jean~Fr{\'e}d{\'e}ric Myoupo.
\newblock Efficient cgm-based parallel algorithms for the longest common
  subsequence problem with multiple substring-exclusion constraints.
\newblock {\em Parallel Computing}, 91:102598, 2020.

\bibitem{ukkonen1985algorithms}
Esko Ukkonen.
\newblock Algorithms for approximate string matching.
\newblock {\em Information and Control}, 64(1-3):100--118, 1985.

\bibitem{xue2015ageseq}
Liang-Jiao Xue and Chung-Jui Tsai.
\newblock Ageseq: analysis of genome editing by sequencing.
\newblock {\em Molecular plant}, 8(9):1428--1430, 2015.

\bibitem{yang2010efficient}
Jiaoyun Yang, Yun Xu, and Yi~Shang.
\newblock An efficient parallel algorithm for longest common subsequence
  problem on {GPU}s.
\newblock In {\em World Congress on Engineering}, volume~1, pages 499--504,
  2010.

\bibitem{zhang2003alignment}
Hongyu Zhang.
\newblock Alignment of blast high-scoring segment pairs based on the longest
  increasing subsequence algorithm.
\newblock {\em Bioinformatics}, 19(11):1391--1396, 2003.

\end{thebibliography}

\clearpage

\iffullversion{
\appendix
\section{More Details for the Landau-Vishkin Algorithm}\label{app:lv-algo}

As mentioned in \cref{sec:intro}, computing the edit distance (the value $G[n,m]$) is equivalent to
finding the shortest distance in a $n\times m$ grid from $(1,1)$ to $(n,m)$.
We will use $x$ and $y$ to denote the row and column ids of a cell, respectively.
Each cell $(x,y)$ in the grid has three out-going edges to its left, right, and bottom-right neighbors (if any).
The edge weight is 0 if it is an edge to the bottom-right and $A[x+1]=B[y+1]$,
and 1 otherwise.
Since the edges have unit weights (or 0), we can use a BFS algorithm to compute the shortest path.
In round $t$, we can process cells with edit distance $t$.
The algorithm terminates when we reach cell $(n,m)$.
As mentioned, the key benefit of using BFS is that not all cells need to be processed.
For example, all cells with $|x-y|>k$ will not be reached when we reach $(n,m)$ with edit distance $k$, since they require at least $k$ edits.
Another important observation is that starting from any cell $(x,y)$,
if there are diagonal edges with weight 0, we should always follow the edges until a unit-weight edge is encountered.
In other words, we should always find the longest common prefix (LCS) $p$ starting from $A[x+1]$ and $B[y+1]$ and skip to the cell at $(x+|p|,y+|p|)$ with no edit.
We show an illustration in \cref{fig:bfs}.
Based on these ideas, it has been proved that a BFS-based edit distance algorithm only needs to visit $O(k^2)$ cells in the DP matrix.

\begin{lemma}
\label{thm:bfs:k2}
  In the BFS-based edit distance algorithm, the total size of the frontiers is $O(k^2)$, where $k$ is the output edit distance.
\end{lemma}

Intuitively, this is because on each diagonal (top-left to bottom-right),
there can be at most one useful cell for each edit distance $t$.
In particular, for multiple cells with the same edit distance, we only need to keep the last one (with the largest row id).
We refer the readers to the previous papers for more analysis of this algorithm~\cite{landau1989fast}. 
This indicates the $O(k^2)$ bound in \cref{thm:bfs:k2}.

The parallel edit distance problem then boils down to designing a parallel BFS algorithm and an efficient data structure to find the LCS
starting from any $A[x]$ and $B[y]$.
We implemented the BFS-based framework in Landau-Vishkin, which is illustrated in \cref{fig:bfs}.
We first use the function \longestprefix{} as a black box. In our paper, we use three implementations for the \longestprefix{}
query, including SA and two hash-based algorithms, introduced in \cref{subsection:algo-BFS}.
As mentioned, we will run the algorithm in rounds, and in round $t$, we will process
a \defn{frontier} of cells, which have edit distance $t$.
Conceptually, the frontier of round $t$
can be obtained from the cells in the frontier of round $t-1$.
Our starting point is cell $(p,p)$, where $p$ is the LCP length of $A$ and $B$, since there is no cost to match the first $p$
characters in $A$ and $B$.
Recall that on each diagonal, we only need to keep the last cell with a certain edit distance.
Therefore, we store the cells in the frontier based on their diagonal ids.
For a cell $(x,y)$, we define its diagonal id as $(x-y)$ (see \cref{fig:bfs}).
Cells with the same $(x-y)$ values are on the same diagonal.
We will represent each frontier $t$ as an array $f_t[\cdot]$,
where $f_t[i]$ represents the last cell in frontier $t$ on diagonal $i$.
Note that we only need to store the x-coordinate in $f_t[i]$, since $y$
can be computed as $x-i$. As mentioned, the starting point is $f_0[0]=LCP(A[1..n], B[1..m])$ (Line \ref{line:bfs-starting} in \cref{algo:bfs_in_code}).

The target cell $(n,m)$ is on diagonal $n-m$ with row id $x=n$, therefore, we start a while-loop as long as the current frontier
has not reached this cell (i.e., $f_t[n-m]\ne n$, Line \ref{line:bfs-outer-loop}).
In the $t$-th round in the while-loop, we generate frontier $t$, which are cells reachable in $t$ edits.
Note that within $t$ edits, the possible diagonal ids $i\in [-t,t]$.
We will enumerate all such diagonal ids $i$, and find the corresponding cell in frontier $t$ on diagonal $i$, i.e., finding $f_t[i]$.
All diagonals can be processed in parallel.
The cell $(x,y)$ in frontier $t$ must be reached from a cell $(x',y')$ in frontier $t-1$.
We use $\langle dx, dy\rangle$ to denote the delta in x- and y-coordinates from $(x',y')$ to $(x,y)$.
Then there are three cases: $\langle 0, 1\rangle$, $\langle 1, 0\rangle$, and $\langle 1,1\rangle$.
To find the cell of $f_t[i]$, we will first track the possible predecessor in each of the three directions,
and compute $f_t[i]$ accordingly.
Note that $x'-y'=(x-dx)-(y-dy)=(x-y)-dx+dy=i-dx+dy$.
Therefore, the diagonal id of $(x',y')$ must be $j=i-dx+dy$ (Line \ref{line:bfs:j}).
If $j$ is within the explored range, we then get $x'=f_{t-1}[j]$.
Accordingly, we can compute the values of $x$ and $y$ (Lines \ref{line:bfs:x}, \ref{line:bfs:y}).
Therefore, we know $(x,y)$ has edit distance $r$.
However, this cell $(x,y)$ is not necessarily the last cell with edit distance $r$ on diagonal $i$.
We will further check the LCP of $A[x+1..n]$ and $B[y+1..m]$, denoting it as $p$.
This means that from $A[x+1]$ and $B[y+1]$, the next $p$ characters are all the same,
and we should match them without any edits.
Therefore, we move the cell to the bottom-right by $p$ cells (Line \ref{line:bfs:addp}).
Finally, for all three directions, we keep the largest $x$ value among them, since we are interested in the last cell
with distance $r$ in each diagonal.
\cref{fig:bfs} shows an example of finding $f_2[0]$, i.e., finding the (last) cell with 2 edits on diagonal 0.

\section{More Details for the Combining Step in the \aalm{} Algorithm}
\label{app:aalm-combine}

\subsection{The Algorithm}\label{app:aalm-combine-algo}
As discussed in \cref{sec:dac-sp}, we need to compute the \SP{} between $v_i\in \dpv_1$ and $u_j\in \dpv_2$
by finding $\min_{l} \dist_1[i,l]+\dist_2[l,j]$,
i.e., for all $w_l$ on the common boundary,
we attempt to use the \SP{} between $v_i$ to $w_l$, and $w_l$ to $u_j$, and find the minimum one (See \cref{fig:dac_mm}(c)).
A na\"ive solution is to use the Floyd-Warshall algorithm~\cite{floyd1962algorithm},
However, it results in $O(n^3)$ work and could be the bottleneck of the algorithm.
The \aalm{} algorithm tackles this challenge by using of the \monge{} property.
We can show the following theorem.

\begin{lemma}[\cite{apostolico1990efficient}]\label{thm:monge}
    Let $\theta(v_i, u_j)$ be the leftmost point on the common boundary
    such that at least one of the shortest paths from $v$ to $u$ goes through it.
    For any $v_{i}$ and $u_{j_1}, u_{j_2}$ s.t., $j_1 < j_2$, $\theta(v_{i}, u_{j_1}) \leq \theta(v_{i}, u_{j_2})$.
\end{lemma}


\begin{figure}[t]
    \centering
    \vspace{-1.0em}
    \includegraphics[width=0.5\columnwidth]{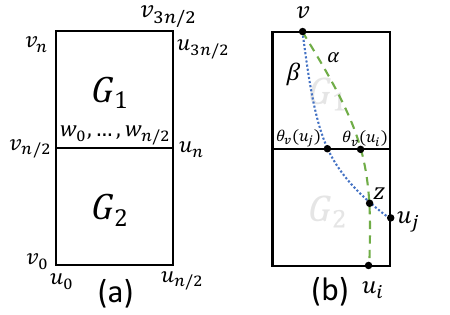}
    \caption{\small\textbf{The illustration of the combining step in the \aalm{} algorithm.} \label{fig:dac_mm_combine}}
    \vspace{-1em}
  \end{figure} 

We call $\theta(v_i,u_j)$ the \defn{best choice} from $v_i$ to $u_j$.
This theorem indicates that the best choices have the monotonic property.
For certain $v_i$, the best choices for $u_1,u_2,\dots,u_n$ is monotonically increasing (going from the left to the right), or more accurately non-decreasing.
The intuition behind the theorem is that two paths that originate from the same point $v$ and
end at two different points $u_i, u_j$ do not cross each other.
If $\theta_v(u_i) > \theta_v(u_j)$, two paths $\alpha, \beta$ will intersect at a point $z$
that does not belong to the boundary of $\dpv_{1\cup 2}$ (See \cref{fig:dac_mm_combine}(b)).
Let $\prefix(\alpha)$ be the length from $v$ to $z$ using the path $\alpha$ (resp., $\prefix(\beta)$).
1) If $\prefix(\alpha) = \prefix(\beta)$, replacing $\alpha$ with $\beta$ yields another shortest path from $v$ to $u_i$,
which crosses the common boundary through a point on the left of $\theta_v(u_i)$, contradicting the definition of $\theta$.
2) If $\prefix(\alpha) \neq \prefix(\beta)$, WLOG, assume the length of $\prefix(\alpha)$ is smaller than $\prefix(\beta)$,
then for the path from $v$ to $u_j$, replacing $\beta$ with $\alpha$ yields a shorter path (contradiction).

With this property, we can combine the \SP{} matrices of $\dpv_1$ and $\dpv_2$ (i.e., $\dist_1$ and $\dist_2$) efficiently using divide-and-conquer.

\begin{lemma}[\cite{apostolico1990efficient}]
    $\dist_1$ and $\dist_2$ can be combined in $O(n^2)$ work and $O(\log^2 n)$ span.
\end{lemma}

Let $X$ be the $(n+1) \times (n/2+1)$ submatrix containing the shortest distance from the top/left
boundaries (i.e., $n/2 \leq v \leq 3n/2$ in \cref{fig:dac_mm_combine}(a))
to the bottom boundary (i.e., $0 \leq w \leq n/2$ in \cref{fig:dac_mm_combine}(a)) of $\dpv_1$.
Similarly, Let $Y$ be the $(n/2+1) \times (n+1)$ submatrix containing the shortest distance from the top
boundary (i.e., $1 \leq w \leq n/2$ in \cref{fig:dac_mm} (b))
to the bottom/right boundaries (i.e., $0 \leq u \leq n$ in \cref{fig:dac_mm} (b)) of $\dpv_2$.
Combining $\dpv_1$ and $\dpv_2$ can be boiled down to computing the product $XY$ in the closed semiring (min, +).
This can be computed recursively:
\begin{enumerate}
    \item[1)] Recursively solve the problem for the product $X'Y'$ where $X'$ (resp., $Y'$) is a $n/2\times n/2$ matrix
     consisting of the even rows (resp., columns) in $X$ (resp., $Y$). This gives $\theta(v, u)$ for all pairs $(v, u)$ whose respective parties are (even, even).
    \item[2)] Compute $\theta(v, u)$ for all pairs $(v, u)$ of parities (even, odd).
    As we already have the (even, even) pairs, we can compute $\theta(v_{2k}, u_{2h+1})$ by only considering the relevant range of the common boundary
     from $\theta(v_{2k-1}, u_{2h+1})$ to $\theta(v_{2k+1}, u_{2h+1})$ according to \cref{thm:monge}.
    \item[3)] Compute $\theta(v, u)$ for all pairs $(v, u)$ of parities (odd, even) using the same method as in 2).
    \item[4)] Compute $\theta(v, u)$ for all pairs $(v, u)$ of parities (odd, odd) using a similar method as in 2).
\end{enumerate}

We note that the ``even/odd'' definition may have minor differences when the array is stored as 0-based or 1-based, but the high-level idea is the same.

This combining step yields recurrences of $W(n, m) = W(n/2, m) + O(nm)$ and $D(n, m)=D(n/2, m)+O((\log n + \log m)^2)$, which gives $W(n) = O(n^2)$ and $D(n)=D(\log^3n)$.

\begin{theorem}[\cite{apostolico1990efficient}]
    The \aalm{} algorithm computes the edit distance in $O(n^2\log n)$ work and $O(\log^3 n)$ span.
\end{theorem}

\subsection{Our Implementation}\label{app:aalm-combine-impl}
The idea behind the aforementioned recursive method is not too complicated, but it is impractical since
it requires allocating and initializing an \SP{} matrix in each recursive call.
In particular, a faithful implementation of this divide-and-conquer \fname{Combine} algorithm
solves a subproblem on all odd-odd pairs (and other parity combinations),
which requires initializing an \SP{} array for such subproblems if we solve them recursively.
However, allocating memory dynamically in parallel is very inefficient, or even impractical.
As a result, we adapt this recursive method to an iterative one to avoid parallel memory allocation.
As mentioned in \cref{sec:dac-sp}, WLOG assume $n=2^p$. In the case where $n$ is not a power of 2,
we just treat $n$ as $n'=2^{\lceil \log n \rceil}$.
Let $\gamma=\log n$.
We will compute the all pair shortest distances between $u_{1..n}$ and $v_{1..n}$ in $\gamma=\log n$ rounds.
In round $r$ ($1\le r\le \gamma$), 
we compute all pair shortest distance between $u_{s\cdot i}$ and $v_{s\cdot j}$ for all possible $i$ and $j$, where $s=2^{\gamma-r}$,
i.e., we select every $s$ elements in $u$, and every $s$ elements in $v$,
and compute the all pair shortest distance between them.
When we finish round $r=\gamma=\log n$, $s=2^0=1$, which means that all pairs between $u$ and $v$ will be processed and we finish computing the \SP{} matrix.
This process simulates the recursive solution in \aalm{}, but can be processed iteratively.

In particular, in round $r$, among all pairs $u_{s\cdot i}$ and $v_{s\cdot j}$,
we already know the \SP{} and best choices for all of even values $s$, i.e., for any $i=2p$ and $j=2q$, because they are processed in the previous round $r-1$.
Therefore, using the best choices of all even values,
for each (odd, even) pair $i=2p+1$, and $j=2q$,
we can narrow down the search for the best choice in a subrange of the common boundary between $\theta(v_{s\cdot 2p}, u_{s\cdot 2q})$
and $\theta(v_{s\cdot (2p+2)}, u_{s\cdot 2q})$.
In this way, we can use a non-recursive algorithm to find the all-pair shortest paths between $v_{i}$ and $u_{j}$,
and thus we only need to use one \SP{} matrix for the entire process, avoiding allocating and initializing an \SP{} matrix in every recursive call.

This algorithm has the same computation as the recursive version, and thus has the same work and span bounds.

\hide{
Let $P_{\odd}: \{i \mid 0 \leq i=(2\times k + 1)p < n \}$.
and $P_{\even}: \{i \mid 0 \leq i=(2\times k)p < n \}$ ($Q_{\odd}$ and $Q_{\even}$ respectively).
We first compute the $\theta(u,v)$ pairs s.t. $u \in P_{\odd}, v \in Q_{odd}$.
Then we can compute the $\theta(u,v)$ pairs s.t. $u \in P_{\odd}, v \in Q_{even}$ and $u \in P_{\even}, v \in Q_{odd}$
by squeezing the range of the common boundary.
After this iteration, we divide both $p$ and $q$ by half, or keep the same value if it is one.
Since all $(u,v)$ pairs s.t. $u \in P_{\even}, v \in Q_{\even}$ in every iteration are the $(u',v')$ pairs s.t. $u \in P_{\odd}$
and $v \in P_{\odd}$ in the last round, they are already computed can be simply skipped.
We can apply the above operations iteratively until $p=1$ and $q=1$,
at which point all the distance pairs are computed.
}

\begin{table}[htbp]
  \small
  \centering
\begin{tabular}{@{}r@{ }@{ }@{ }l|r@{}r@{}r|r@{}r@{}r|r@{}r@{}r|r|r}
  \multicolumn{1}{c}{\multirow{2}[1]{*}{\boldmath{}\textbf{$\boldsymbol{n}$}\unboldmath{}}} & \multicolumn{1}{c|}{\multirow{2}[1]{*}{\boldmath{}\textbf{$\boldsymbol{k}$}\unboldmath{}}} & \multicolumn{3}{c|}{\textbf{\bfshash{}}} & \multicolumn{3}{c|}{\textbf{\bfssimplehash{}}} & \multicolumn{3}{c|}{\textbf{\bfssa{}}} & \multicolumn{1}{c|}{\multirow{2}[1]{*}{\textbf{DaC}}} & \multicolumn{1}{c}{\multirow{2}[1]{*}{\textbf{Parlay}}} \\
        &       & \multicolumn{1}{@{ }c@{ }}{\textbf{Build}} & \multicolumn{1}{@{ }c@{ }}{\textbf{Query}} & \multicolumn{1}{@{ }c|}{\textbf{Total}} & \multicolumn{1}{@{ }c@{ }}{\textbf{Build}} & \multicolumn{1}{@{ }c@{ }}{\textbf{Query}} & \multicolumn{1}{@{ }c@{ }|}{\textbf{Total}} & \multicolumn{1}{@{ }c@{ }}{\textbf{Build}} & \multicolumn{1}{@{ }c@{ }}{\textbf{Query}} & \multicolumn{1}{@{ }c@{ }|}{\textbf{Total}} &       &  \\
  \midrule
  \boldmath{}\textbf{$10^6$}\unboldmath{} & \boldmath{}\textbf{$1$}\unboldmath{} & 0.001 & \textless.001 & \underline{\textless.001} &0.001 & \textless.001 & \textless.001 & 0.102 & \textless.001 & 0.102 & 0.033 & 14.4 \\
  \boldmath{}\textbf{$10^6$}\unboldmath{} & \boldmath{}\textbf{$10$}\unboldmath{} & 0.001 & \textless.001 & \underline{\textless.001} & 0.001 & \textless.001 & \textless.001 & 0.101 & \textless.001 & 0.101 & 0.126 & 14.6 \\
  \boldmath{}\textbf{$10^6$}\unboldmath{} & \boldmath{}\textbf{$10^2$}\unboldmath{} & 0.001 & 0.002 & 0.003 & 0.001 & \textless.001 & \underline{0.002} & 0.095 & 0.001 & 0.096 & 7.65  & 14.6 \\
  \boldmath{}\textbf{$10^6$}\unboldmath{} & \boldmath{}\textbf{$10^3$}\unboldmath{} & 0.001 & 0.032 & 0.033 & 0.001 & 0.021 & \underline{0.022} & 0.087 & 0.021 & 0.108 & 242   & 14.6 \\
  \boldmath{}\textbf{$10^6$}\unboldmath{} & \boldmath{}\textbf{$10^4$}\unboldmath{} & 0.001 & 0.427 & 0.428 & 0.001 & 0.361 & \underline{0.362} & 0.072 & 0.368 & 0.440 & t.o.  & 14.5 \\
  \boldmath{}\textbf{$10^6$}\unboldmath{} & \boldmath{}\textbf{$10^5$}\unboldmath{} & 0.001 & 8.56  & 8.56  & 0.009 & 6.78  & \underline{6.79} & 0.046 & 7.23  & 7.28  & t.o.  & 14.1 \\
  \midrule
  \boldmath{}\textbf{$10^7$}\unboldmath{} & \boldmath{}\textbf{$1$}\unboldmath{} & 0.005 & \textless.001 & \underline{0.005} & 0.007 & \textless.001 & 0.007 & 0.765  & 0.002 & 0.767  & 0.203 & t.o. \\
  \boldmath{}\textbf{$10^7$}\unboldmath{} & \boldmath{}\textbf{$10$}\unboldmath{} & 0.004 & \textless.001 & \underline{0.004} & 0.009 & \textless.001 & 0.009 & 0.738  & \textless.001 & 0.740  & 1.60  & t.o. \\
  \boldmath{}\textbf{$10^7$}\unboldmath{} & \boldmath{}\textbf{$10^2$}\unboldmath{} & 0.003 & 0.004 & \underline{0.007} & 0.006 & \textless.001 & 0.007 & 0.667  & 0.002 & 0.669  & 58.6  & t.o. \\
  \boldmath{}\textbf{$10^7$}\unboldmath{} & \boldmath{}\textbf{$10^3$}\unboldmath{} & 0.003 & 0.040 & 0.043 & 0.006 & 0.022 & \underline{0.028} & 0.599  & 0.022 & 0.621  & t.o.  & t.o. \\
  \boldmath{}\textbf{$10^7$}\unboldmath{} & \boldmath{}\textbf{$10^4$}\unboldmath{} & 0.002 & 0.508 & 0.510 & 0.007 & 0.383 & \underline{0.391} & 0.503  & 0.382 & 0.885  & t.o.  & t.o. \\
  \boldmath{}\textbf{$10^7$}\unboldmath{} & \boldmath{}\textbf{$10^5$}\unboldmath{} & 0.002 & 10.5  & 10.5  & 0.010 & 8.02  & \underline{8.03} & 0.376 & 8.68  & 9.05   & t.o.  & t.o. \\
  \midrule
  \boldmath{}\textbf{$10^8$}\unboldmath{} & \boldmath{}\textbf{$1$}\unboldmath{} & 0.022 & \textless.001 & \underline{0.022} & 0.052 & \textless.001 & 0.052 & 9.46   & 0.040 & 9.50   & 2.83  & t.o. \\
  \boldmath{}\textbf{$10^8$}\unboldmath{} & \boldmath{}\textbf{$10$}\unboldmath{} & 0.027 & \textless.001 & \underline{0.028} & 0.053 & \textless.001 & 0.054 & 9.04   & 0.036 & 9.07   & 39.3  & t.o. \\
  \boldmath{}\textbf{$10^8$}\unboldmath{} & \boldmath{}\textbf{$10^2$}\unboldmath{} & 0.021 & 0.005 & \underline{0.026} & 0.050 & 0.002 & 0.052 & 8.56   & 0.036 & 8.60   & t.o.  & t.o. \\
  \boldmath{}\textbf{$10^8$}\unboldmath{} & \boldmath{}\textbf{$10^3$}\unboldmath{} & 0.024 & 0.053 & \underline{0.077} & 0.051 & 0.028 & 0.079 & 8.40   & 0.058 & 8.46   & t.o.  & t.o. \\
  \boldmath{}\textbf{$10^8$}\unboldmath{} & \boldmath{}\textbf{$10^4$}\unboldmath{} & 0.020 & 0.605 & 0.625 & 0.053 & 0.486 & \underline{0.538} & 7.59   & 0.412 & 8.00   & t.o.  & t.o. \\
  \boldmath{}\textbf{$10^8$}\unboldmath{} & \boldmath{}\textbf{$10^5$}\unboldmath{} & 0.019 & 11.8  & 11.8  & 0.074 & 9.38   & \underline{9.46} & 6.61   & 9.92   & 16.5  & t.o.  & t.o. \\
  \midrule
  \boldmath{}\textbf{$10^9$}\unboldmath{} & \boldmath{}\textbf{$1$}\unboldmath{} & 0.200 & \textless.001 & \underline{0.200} & 0.501  & \textless.001 & 0.500  & 122   & 0.526 & 123   & 40.2  & t.o. \\
  \boldmath{}\textbf{$10^9$}\unboldmath{} & \boldmath{}\textbf{$10$}\unboldmath{} & 0.203 & \textless.001 & \underline{0.204} & 0.500  & \textless.001 & 0.501  & 117   & 0.554 & 118   & t.o.  & t.o. \\
  \boldmath{}\textbf{$10^9$}\unboldmath{} & \boldmath{}\textbf{$10^2$}\unboldmath{} & 0.207 & 0.007 & \underline{0.214} & 0.498  & 0.003 & 0.501  & 111   & 0.489 & 112   & t.o.  & t.o. \\
  \boldmath{}\textbf{$10^9$}\unboldmath{} & \boldmath{}\textbf{$10^3$}\unboldmath{} & 0.200 & 0.068 & \underline{0.268} & 0.507  & 0.035 & 0.543  & 105   & 0.722 & 106   & t.o.  & t.o. \\
  \boldmath{}\textbf{$10^9$}\unboldmath{} & \boldmath{}\textbf{$10^4$}\unboldmath{} & 0.201 & 0.777 & \underline{0.978} & 0.499  & 0.517 & 1.02  & 97.7    & 0.895 & 98.6    & t.o.  & t.o. \\
  \boldmath{}\textbf{$10^9$}\unboldmath{} & \boldmath{}\textbf{$10^5$}\unboldmath{} & 0.197 & 14.5  & 14.6  & 0.583  & 11.4  & \underline{12.0} & 90.2    & 12.1  & 102   & t.o.  & t.o. \\
  \end{tabular}%

  \caption{\small\textbf{Full experimental result of the running time on synthetic datasets (in seconds). }
  Smaller is better. ``t.o.'' $=$ timeout.
  The fastest algorithm on each row is underlined.
  ``Build'' = Building time.
  ``DaC'' = Our \dacmm{} algorithm.
  ``Parlay'' = ParlayLib~\cite{blelloch2020parlaylib}.
  }
  \label{tab:full}
  \end{table}%
 
\begin{table}[htbp]
  \small
  \centering
  \caption{\small\textbf{Running time of our \bfssa{} implementation and the \bfssa{} implementation in ParlayLib~\cite{blelloch2020parlaylib} (in seconds).}
  Smaller is better.
  The fastest algorithm on each row is underlined.}
    \begin{tabular}{ll|rrr|rrr}
    \multicolumn{1}{c}{\multirow{2}[1]{*}{\boldmath{}\textbf{$\boldsymbol{n}$}\unboldmath{}}} & \multicolumn{1}{c|}{\multirow{2}[1]{*}{\boldmath{}\textbf{$\boldsymbol{k}$}\unboldmath{}}} & \multicolumn{3}{c|}{\textbf{\bfssa{}}} & \multicolumn{3}{c}{\textbf{BFS-SA (ParlayLib)}} \\
          &       & \multicolumn{1}{c}{\textbf{Building}} & \multicolumn{1}{c}{\textbf{Query}} & \multicolumn{1}{c|}{\textbf{Total}} & \multicolumn{1}{c}{\textbf{Building}} & \multicolumn{1}{c}{\textbf{Query}} & \multicolumn{1}{c}{\textbf{Total}} \\
    \midrule
    \boldmath{}\textbf{$10^6$}\unboldmath{} & \boldmath{}\textbf{$1$}\unboldmath{} & 0.102 & \textless.001 & \underline{0.102} & 0.235  & \textless.001  & 0.236 \\
    \boldmath{}\textbf{$10^6$}\unboldmath{} & \boldmath{}\textbf{$10$}\unboldmath{} & 0.101 & \textless.001 & \underline{0.101} & 0.216  & \textless.001  & 0.216 \\
    \boldmath{}\textbf{$10^6$}\unboldmath{} & \boldmath{}\textbf{$10^2$}\unboldmath{} & 0.095 & \textless.001 & \underline{0.096} & 0.169  & \textless.001  & 0.170 \\
    \boldmath{}\textbf{$10^6$}\unboldmath{} & \boldmath{}\textbf{$10^3$}\unboldmath{} & 0.087 & 0.021 & \underline{0.108} & 0.126  & 0.022  & 0.147 \\
    \boldmath{}\textbf{$10^6$}\unboldmath{} & \boldmath{}\textbf{$10^4$}\unboldmath{} & 0.072 & 0.348 & \underline{0.420}  & 0.068  & 0.362  & 0.430 \\
    \boldmath{}\textbf{$10^6$}\unboldmath{} & \boldmath{}\textbf{$10^5$}\unboldmath{} & 0.046 & 7.23  & \underline{7.28}  & 0.032  & 7.45  & 7.48 \\
    \midrule
    \boldmath{}\textbf{$10^7$}\unboldmath{} & \boldmath{}\textbf{$1$}\unboldmath{} & 0.765  & 0.002 & \underline{0.767}  & 2.86  & 0.001  & 2.86 \\
    \boldmath{}\textbf{$10^7$}\unboldmath{} & \boldmath{}\textbf{$10$}\unboldmath{} & 0.739  & 0.001 & \underline{0.740}  & 2.54  & 0.002  & 2.54 \\
    \boldmath{}\textbf{$10^7$}\unboldmath{} & \boldmath{}\textbf{$10^2$}\unboldmath{} & 0.667  & 0.002 & \underline{0.670}  & 2.04  & 0.002  & 2.04 \\
    \boldmath{}\textbf{$10^7$}\unboldmath{} & \boldmath{}\textbf{$10^3$}\unboldmath{} & 0.599   & 0.022 & \underline{0.621}  & 1.58  & 0.022  & 1.61 \\
    \boldmath{}\textbf{$10^7$}\unboldmath{} & \boldmath{}\textbf{$10^4$}\unboldmath{} & 0.503   & 0.382 & \underline{0.885}  & 1.13  & 0.373  & 1.50 \\
    \boldmath{}\textbf{$10^7$}\unboldmath{} & \boldmath{}\textbf{$10^5$}\unboldmath{} & 0.376  & 8.68  & \underline{9.05}   & 0.648  & 9.08  & 9.73 \\
    \midrule
    \boldmath{}\textbf{$10^8$}\unboldmath{} & \boldmath{}\textbf{$1$}\unboldmath{} & 9.46   & 0.040  & \underline{9.50}   & 30.1    & 0.047  & 30.2 \\
    \boldmath{}\textbf{$10^8$}\unboldmath{} & \boldmath{}\textbf{$10$}\unboldmath{} & 9.04     & 0.036 & \underline{9.07}   & 27.3    & 0.036  & 27.4 \\
    \boldmath{}\textbf{$10^8$}\unboldmath{} & \boldmath{}\textbf{$10^2$}\unboldmath{} & 8.56   & 0.036 & \underline{8.60}   & 23.1    & 0.028  & 23.1 \\
    \boldmath{}\textbf{$10^8$}\unboldmath{} & \boldmath{}\textbf{$10^3$}\unboldmath{} & 8.40   & 0.058 & \underline{8.46}   & 19.8    & 0.053  & 19.9 \\
    \boldmath{}\textbf{$10^8$}\unboldmath{} & \boldmath{}\textbf{$10^4$}\unboldmath{} & 7.59   & 0.412 & \underline{8.00}     & 14.5    & 0.425  & 14.9 \\
    \boldmath{}\textbf{$10^8$}\unboldmath{} & \boldmath{}\textbf{$10^5$}\unboldmath{} & 6.61   & 9.92   & \underline{16.5}  & 9.95    & 10.0    & 20.0 \\
    \midrule
    \boldmath{}\textbf{$10^9$}\unboldmath{} & \boldmath{}\textbf{$1$}\unboldmath{} & 122   & 0.526 & \underline{123}   & 357   & 0.528  & 358 \\
    \boldmath{}\textbf{$10^9$}\unboldmath{} & \boldmath{}\textbf{$10$}\unboldmath{} & 117   & 0.554 & \underline{118}   & 310   & 0.468  & 311 \\
    \boldmath{}\textbf{$10^9$}\unboldmath{} & \boldmath{}\textbf{$10^2$}\unboldmath{} & 111   & 0.489 & \underline{112}  & 250   & 0.472  & 250 \\
    \boldmath{}\textbf{$10^9$}\unboldmath{} & \boldmath{}\textbf{$10^3$}\unboldmath{} & 105   & 0.722 & \underline{106}   & 209   & 0.550  & 210 \\
    \boldmath{}\textbf{$10^9$}\unboldmath{} & \boldmath{}\textbf{$10^4$}\unboldmath{} & 97.7    & 0.895 & \underline{98.6}    & 160   & 0.930  & 161 \\
    \boldmath{}\textbf{$10^9$}\unboldmath{} & \boldmath{}\textbf{$10^5$}\unboldmath{} & 90.2    & 12.1  & \underline{102}   & 115   & 12.2    & 127 \\
    \end{tabular}%
  \label{tab:sa_compare}%
\end{table}%

}

\end{document}